\documentclass{article}
\usepackage[a4paper, total={6in, 8in}, margin=1.2in]{geometry}
\usepackage[utf8]{inputenc}
\usepackage[table,xcdraw]{xcolor}
\usepackage{authblk}
\usepackage{graphicx}
\usepackage{newtxtext}
\usepackage{newtxmath}
\usepackage{hyperref}
\usepackage{subcaption}
\hypersetup{
	colorlinks = true,
	urlcolor   = blue,
	citecolor  = black,
}

\newcommand{\RomanNumeralCaps}[1]
\linenumbers

\newcommand{\revOne}[1]{#1}
\newcommand{\revTwo}[1]{#1}
\newcommand{\revThree}[1]{#1}
\newcommand{\revIntern}[1]{#1}

\title{Linear and non-linear receptivity mechanisms in boundary layers subject to free-stream turbulence}

\author[1]{Diego C. P. Blanco \thanks{Correspondence address: diegodcpb@ita.br}}
\author[2]{Ardeshir Hanifi}
\author[2]{Dans S. Henningson}
\author[1]{André V. G. Cavalieri}

\affil[1]{Divisão de Engenharia Aeroespacial, Instituto Tecnológico de Aeronáutica, 12228-900, São José dos Campos/SP - Brazil}
\affil[2]{KTH Engineering Mechanics, FLOW Turbulence Lab, KTH Royal Institute of Technology, SE-10044 Stockholm, Sweden}

\date{}

\begin{document}
\maketitle

\begin{abstract}
Large-eddy simulations of a flat-plate boundary layer, without a leading edge, subject to multiple levels of incoming free stream turbulence are considered in the present work. Within an input-output model where non-linear terms of the incompressible Navier-Stokes equations are treated as an external forcing, we manage to separate inputs related to perturbations coming through the intake of the numerical domain, whose evolution represents a linear mechanism, and the volumetric non-linear forcing due to triadic interactions. With these, we perform the full reconstruction of the statistics of the flow, as measured in the simulations, to quantify pairs of wavenumbers and frequencies more affected by either linear or non-linear receptivity mechanisms. Inside the boundary layer, different wavenumbers at near-zero frequency reveal streaky structures. Those that are amplified predominantly via linear interactions with the incoming vorticity occur upstream and display transient growth, while those generated by the non-linear forcing are the most energetic and appear in more downstream positions. The latter feature vortices growing proportionally to the laminar boundary layer thickness, along with a velocity profile that agrees with the optimal amplification obtained by linear transient growth theory. The numerical approach presented is general and could potentially be extended to any simulation for which receptivity to incoming perturbations needs to be assessed.
\end{abstract}

\section{Introduction} \label{sec:intro}

\revOne{Boundary-layer} flows are among the most studied problems in fluid dynamics due to their practical importance in the determination of \revOne{skin-friction} drag of objects, heat transfer and stall characteristics in airplane wings and turbine blades.

Nevertheless, to this day, there is no general mathematical model capable of predicting the transition from laminar to turbulent flow under all conditions, even for the simplest case of a boundary layer over a flat plate without pressure gradient \cite{doi:10.1146/annurev.fluid.34.082701.161921,fransson_shahinfar_2020}. This unpredictability is mainly due to the multiple parameters that are known to affect transition, such as \revOne{free-stream} turbulence intensity, sound, surface roughness, \revOne{leading-edge} shape, and the still incomplete knowledge of how these parameters interact.

Concerning environmental effects, a simplified roadmap to turbulence is described by \cite{morkovin1994} as a function of disturbance amplitudes, with transition beginning with the process denoted receptivity \cite{morkovin1969}, in which wave-like disturbances originating in the free flow enter the boundary layer.

If the magnitude of environmental disturbances is weak, the initial growth of boundary layer instabilities can be described by modal stability theory, which predicts the exponential evolution of the primary unstable modes (eigenfunctions) of the Orr-Sommerfeld-Squire (OSS) equations over relatively long lengths \cite{Reed1996}. In boundary layer over flat plates, subject to no pressure gradient, these primary instabilities are two-dimensional oscillatory modes called Tollmien–Schlichting (TS) waves \cite{SCHUBAUER1947}. Then, at large enough perturbation amplitudes, non-linear effects take place and the unstable linear modes lose symmetry, degenerating in secondary instabilities before breaking into turbulent spots due to non-linear mechanisms.

On the other hand, in the presence of stronger environmental forcing, turbulent spots inside the boundary layer appear much sooner than predicted by modal stability, completely bypassing primary mode growth. This phenomenon, therefore called bypass transition \cite{morkovin1969,morkovin1985}, has since been associated with cases such as rough surfaces \cite{reshotko1984disturbances,Morkovin1990,denissen2008roughness,Deyn2020} and high free-stream turbulence levels, above around 1\% \cite{morkovin1985,suder1988experimental,matsubara2001disturbance},  where linear theory predictions fail and receptivity mechanisms are still not well understood.

Initially, bypass transition was thought to be mainly a result of non-linear phenomena, a notion that was later challenged by the concept of transient growth \cite{Reshotko2001}, developed in the early 1990s and formalised in \cite{Schmid1993}. Due to the non-orthogonality of the OSS operator, the superposition of eigenfunctions can lead to a transient algebraic growth, even in cases where the boundary layer is linearly stable, i.e., below the critical Reynolds number for the occurrence of TS waves.

Transient growth theory, often referred to as non-modal stability theory, is based on the lift-up effect first demonstrated by \cite{ellingsen1975stability} and later developed by \cite{landahl_1980}, where three-dimensional infinitesimal disturbances can grow algebraically in parallel inviscid shear flows, regardless of the modal stability conditions. Moreover, \cite{landahl_1980} formally connected this behaviour with the low frequency longitudinal streaky structures first identified in transitional and turbulent boundary layers by \cite{klebanoff1971effect}, later found to be important in all transitional and turbulent shear flows \cite{brandt2014}. The magnitude of the transient growth is an important parameter that defines the path to turbulence. Weaker streaks may simply decay, giving space to primary mode growth, or lead to secondary instabilities. Stronger streaks, however, might degenerate directly into turbulent spots.

In the specific case of bypass transition in boundary layers due to FST, two distinct receptivity mechanisms have been proposed \cite{brandt_schlatter_henningson_2004}: a linear mechanism caused by perturbations at the leading edge and a non-linear one, caused by interactions between oblique waves above the boundary layer.

When vortical disturbances are present at the leading edge, low-frequency perturbations induce streamwise vortices of alternating direction that, in turn, cause the linear transient growth of streaky structures inside the laminar boundary layer \cite{butler1992,doi:10.1063/1.869908,luchini2000}. These streaks are characterised by alternating regions of fast and slow longitudinal flow. In locations where the streamwise vortices carry matter downwards to the wall, a fast (positive) streak is generated, while the outflow from the wall generates slow (negative) streaks. The profiles for the optimal response of streaks induced by this mechanism consistently match experiments \cite{Kendall,matsubara2001disturbance}, as discussed by \cite{luchini2000}.

On the other hand, when disturbances are found above the boundary layer, the transition can be triggered by pairs of oblique waves propagating at the same frequency, $\omega$, and opposite spanwise wavenumbers, $\pm\beta$, generating structures in the boundary layer, through quadratic non-linear interactions, which are associated with double the initial wavenumber, i.e. $(\pm \beta,\omega) \rightarrow (2\beta,0)$. This mechanism is also known to generate streamwise vortices and streaks \cite{Schmid1996}, a process verified both numerically and experimentally \cite{berlin_wiegel_henningson_1999} and modelled via weakly non-linear analysis \cite{doi:10.1063/1.1456062}.

In this work, a set of numerical simulations of a boundary layer subject to different levels of free-stream turbulence (FST) is considered, to study in detail the process of receptivity to external vorticity. Modal analysis, namely spectral POD \cite{towne_schmidt_colonius_2018}, and resolvent analysis \cite{jovanovic_bamieh_2005,mckeon_sharma_2010} are employed,
in combination with the ideas developed in \cite{morra_nogueira_cavalieri_henningson_2021} and \cite{nogueira_morra_martini_cavalieri_henningson_2021} \revThree{which, in turn, arise from the realisation that accurate predictions from linear models require accurate knowledge of the non-linear forcing statistics \cite{chevalier_hœpffner_bewley_henningson_2006}, which would otherwise be modelled as incoherent (white) noise \cite{hœpffner_chevalier_bewley_henningson_2005}}. The coloured statistics of the non-linear forcing term are computed directly from the simulated data and, instead of computing spectral POD modes of the forcing, we obtain, via the resolvent-based extended spectral POD method \cite{karban_martini_cavalieri_lesshafft_jordan_2022}, response and forcing modes which are related by the resolvent operator. This setup allows for the identification of coherent structures that are more affected by either linear or non-linear interactions with vortical free stream disturbances of complex nature. In the latter case, the non-linear forcing capable of generating said coherent structures is characterised.

The separated consideration of linear and non-linear mechanisms in the resolvent framework allows exploring how streaks present in the data can be connected to upstream disturbances through a linear receptivity, or to triadic interactions in a non-linear receptivity. The dominance of each mechanism in different regions of the boundary layer may thus be quantified using simulation data.

This manuscript is divided in the following manner: section \ref{sec:simulations} describes in detail the numerical setup employed in the study; section \ref{sec:techniques} exposes the mathematical formulation capable of separating linear and non-linear receptivity mechanisms and the spectral analysis procedures; sections \ref{sec:spectra} to \ref{sec:mechs} present and discuss the results, discriminating the differences found between linearly and non-linearly induced structures inside and outside the boundary layer. The manuscript is completed with conclusions in section \ref{sec:conclusions}.

\section{Boundary layer simulations} \label{sec:simulations}

We performed multiple simulations of boundary-layer flows subject to different levels of free-stream turbulence (FST) varying from $Tu=0.5\%$ to $3.5\%$, in steps of $0.5\%$, in a total of 7 different cases. The databases were obtained using the SIMSON pseudo-spectral solver \cite{chevalier2007simson}. These are large-eddy simulations (LES) of transitional regimes in a Blasius-type boundary layer over a flat plate without a leading edge and zero pressure gradient, performed using an approximate deconvolution model with relaxation terms (ADM-RT) \cite{doi:10.1063/1.1350896,doi:10.1080/14685240600602929}.

\subsection{Numerical setup}

Each simulation was set according to \cite{sasaki_morra_cavalieri_hanifi_henningson_2020}, based on the work of \cite{brandt_schlatter_henningson_2004}, and consists of a $231 \times 121 \times 108$ $(x \times y \times z)$ \revTwo{Cartesian} grid constructed with Chebyshev nodes in the $y$ direction, perpendicular to the wall, and homogeneously spaced nodes in the streamwise and spanwise directions, $x$ and $z$. The boundary layer is started with a finite thickness. \revTwo{All variables are non-dimensionalised by the reference length $\delta^*_0$, the boundary-layer displacement thickness at the intake, and a time scale $t = \delta^*_0/U_\infty$, where $U_\infty$ is the free-stream velocity.} The numerical domain is a box of size $x \in [0,1000]$, $y \in [0,60]$ and $z \in [-25,25]$. Both $x$ and $z$ directions are periodic and decomposed in Fourier modes, while the $y$ direction uses a Chebyshev polynomial basis. Periodicity in the streamwise direction is achieved by the introduction of a fringe region contained in the range $x \in [910,1000]$.

\begin{figure}
	\centering
	\includegraphics[width=\linewidth]{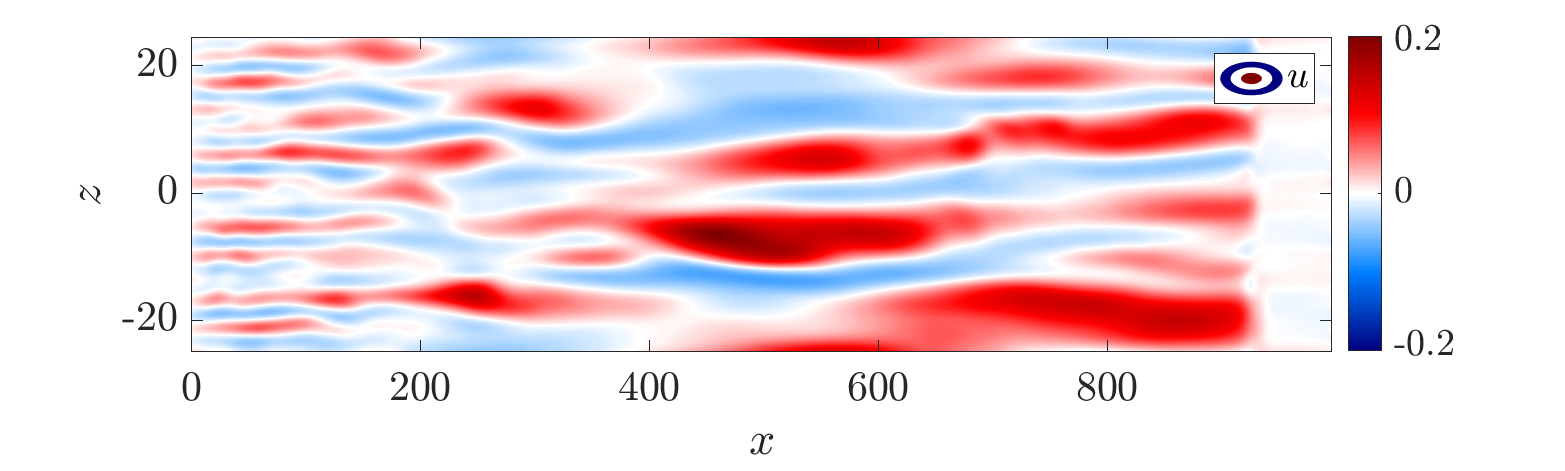}
	\caption{Snapshot of a simulation with $Tu=3.5\%$ incoming FST level, where streaky structures can be identified. Slice at $y = 0.8$, inside the boundary layer.}
	\label{fig:snapshot}
\end{figure}

 At the intake, $Re_{\delta_0^*}= U_\infty \delta^*_0/\nu = 300$, with $\nu$ being the kinematic viscosity of the fluid. At this Reynolds number, the boundary layer is linearly stable and, thus, Tollmien-Schlichting (TS) waves are not expected to be significant over the relatively short extent of the domain. Instead, streamwise elongated (streaky) structures are observed in the simulations at the highest FST levels investigated, as shown in figure \ref{fig:snapshot}.

\begin{figure}
	\centering
	\includegraphics[width=0.9\linewidth]{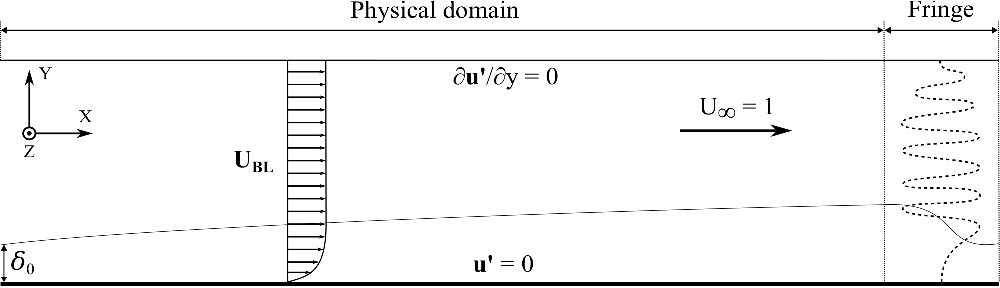}
	\caption{Diagram of the boundary-layer setup, showing boundary conditions. $x$ and $z$ directions are periodic and $\delta_0$ is the initial boundary layer thickness. $\mathbf{U_{BL}}(x,y)$ is the Blasius base flow and $\mathbf{u^\prime}(x,y,z)$ are velocity fluctuations. Legend: (dotted line) Forcing modes from the continuous branch of the Orr-Sommerfeld-Squire (OSS) operator.}
	\label{fig:sketchbl}
\end{figure}

The no-slip condition
\begin{equation}\label{eq:noslip}
	\mathbf{u^\prime}(x,0,z,t) = 0
\end{equation}
is imposed on the wall and the Neumann condition
\begin{equation}\label{eq:newmann_bc}
	\frac{\partial}{\partial y} \mathbf{u^\prime}(x,60,z,t) = 0
\end{equation}
is applied on the upper boundary, with $\mathbf{u^\prime}(x,y,z,t)$ representing velocity fluctuations with respect to the 2D Blasius base flow, $\mathbf{U}_{BL}(x,y)$ (figure \ref{fig:sketchbl}). For all performed simulations, the physical domain ends before the development of turbulent spots in the boundary layer, i.e., before the transition to the turbulent regime. The use of a short spatial domain reduces the computational cost of the present study, which involves detailed post-processing of several numerical simulations. Moreover, by restricting the domain to the initial development of streaks we can focus on the receptivity stage, before the actual transition to turbulence that would occur at larger values of $x$.

Concerning the time evolution, linear terms of the Navier-Stokes (NS) equations are implicitly marched with a second-order Crank-Nicolson scheme, while an explicit third-order, four-stage, Runge-Kutta scheme is applied over non-linear terms. For each simulation, we compute a total of 2000 snapshots, taken in time steps of $\Delta t=10$, of fully developed, statistically stationary, flow.

\subsection{Fringe region forcing} \label{sec:fringe_forcing}

Some assumptions are made to synthesize valid inflow conditions at $x=0$ and circumvent the need to compute a turbulent field upstream of the flat plate or the flow around a leading edge. Isotropic and homogeneous FST is introduced in the simulations by forcing several modes in the continuous branch of the linearised Orr-Sommerfeld-Squire (OSS) operator within the fringe region, as illustrated in figure \ref{fig:sketchbl}.

\begin{figure}
	\centering
	\includegraphics[width=0.5\linewidth]{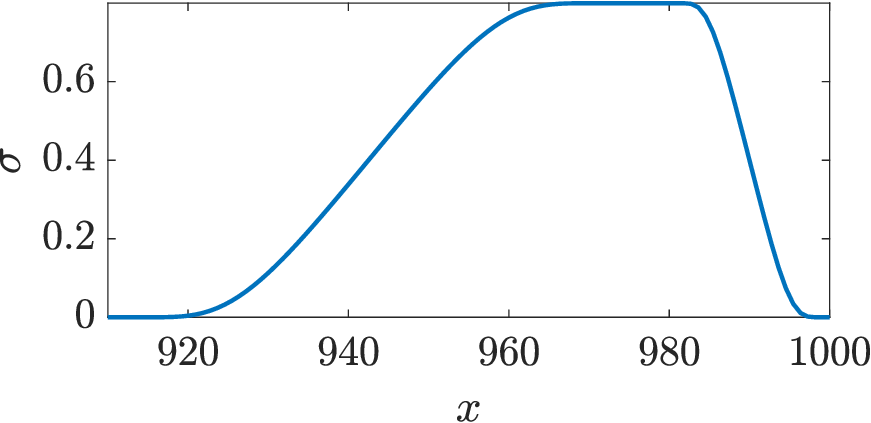}
	\caption{Fringe gain function, \revTwo{the same described in \cite{chevalier2007simson}}. The maximum gain inside the fringe is set to 0.8.}
	\label{fig:gain_fringe}
\end{figure}

The FST generation procedure is referred to in \cite{Schlatter2001DirectNS} and \cite{brandt_schlatter_henningson_2004}, based on the methods presented in \cite{grosch_salwen_1978} and \cite{jacobs_durbin_2001}. Considering the linearised Navier-Stokes (LNS) momentum equations in perturbation form around a base flow \revOne{and non-linear fluctuations terms gathered into the function $f(\mathbf{u^\prime})$}, we force a desired velocity vector $\mathbf{\zeta}(x,y,z,t)$ inside the fringe following the formulation
\begin{equation}\label{eq:ns_1}
	\left.
	\begin{aligned}
		\frac{\partial \mathbf{u^\prime}}{\partial t} = LNS (\mathbf{u^\prime},\mathbf{U_{BL}}) + \revOne{f(\mathbf{u^\prime})} + \sigma(x)(\mathbf{\zeta}-\mathbf{u^\prime}),& \\
		\nabla \cdot \mathbf{u^\prime} = 0,&
	\end{aligned}
	\right\}
\end{equation}
where $\sigma(x)$ is a gain function, which is positive inside the fringe region and null everywhere else (figure \ref{fig:gain_fringe}). The term $\sigma(x)(\mathbf{\zeta}-\mathbf{u^\prime})$ is thus responsible for smoothly changing the fluctuation field entering the left side of the fringe region towards the desired reference forcing vector $\mathbf{\zeta}$ introduced in the fringe.

Isotropic homogeneous turbulence can be represented as a sum of Fourier modes with random amplitude \cite{rogallo1981numerical}. In the boundary layer case, however, this approach is not capable of modelling the presence of the wall, as the $y$ direction is non-homogeneous. For this application, a basis composed of modes in the continuous spectrum of the OSS operator is assumed to be a reasonable choice to satisfy all the necessary boundary conditions. These modes tend to Fourier modes far from the wall and decay to zero near it, \revTwo{generating a perturbation field mainly localised outside the boundary layer, as shown in Appendix \ref{sec:propPert}}. On the other hand, modes of the discrete spectrum are only significant inside the boundary layer and decay exponentially farther from the wall, not being suitable in this application \cite{grosch_salwen_1978}.

By computing eigenfunctions in the continuous branch of the OSS spectrum, $\mathbf{u^\prime_{OSS}}$, normalized to unit energy, we can write the expansion for an arbitrary perturbation vector
\begin{equation}\label{eq:expansion}
	\mathbf{\zeta}(x,y,z,t) = Re\left\{\sum_{\omega} \sum_{\gamma} \sum_{\beta} \Phi(\omega,\gamma,\beta) \: \mathbf{u^\prime_{OSS}}(\omega,y,\beta) \: e^{i\left(Re\{\alpha(\omega,\gamma,\beta)\} x + \beta z - \omega t \right)}\right\},
\end{equation}
where $\omega$, $\gamma$ and $\beta$ are real parameters and $\alpha(\omega,\gamma,\beta)$ is the complex eigenvalue of $\mathbf{u^\prime_{OSS}}$  computed via spatial stability \cite{jacobs_durbin_2001}. Here, $\alpha,\gamma,\beta$ are respectively the wavenumbers in the $x,y,z$ directions and $\omega$ the frequency. The factor $\Phi$ is the energy scaling applied to match the von Karman spectrum, discussed in the following paragraphs. Only the real part of $\alpha$ is taken inside the exponent to maintain forcing fluctuation at a fixed magnitude throughout the fringe zone's streamwise extension, ignoring in practice the effects of viscous attenuation \cite{brandt_schlatter_henningson_2004}.

We consider wavenumbers $\kappa = \sqrt{Re\{\alpha\}^2+\gamma^2+\beta^2}$ equally spaced within the range limited by the numerical resolution of the simulations, $\kappa \in [\kappa_l,\kappa_u]$. In general, $\kappa_l$ is a function of the domain size, while $\kappa_u$ is bounded by the resolution of the grid. For simplification, we replace $\omega = \alpha U_\infty$, considering that modes of the continuous spectrum have phase speed equal to $U_\infty$, to define a tridimensional space of parameters $(\omega,\gamma,\beta)$ for which a given value $\kappa$ is represented by a spherical shell \cite{brandt_schlatter_henningson_2004}. We select $N_s$ shells, within which we include $N_\kappa$ combinations of the $(\omega,\gamma,\beta)$ parameters of constant $\kappa$, filling the surface with equally spaced points \cite{Schlatter2001DirectNS}. The value $\gamma=0$ is avoided. A random rotation is applied to each shell at every time step to further improve isotropy. In this work, we adopt the values $\kappa_l=0.23$, $\kappa_u=3.0$, $N_s = 20$ and $N_\kappa = 10$, in a total  of $N_s N_\kappa = 200$ eigenfunctions, the same as in \cite{sasaki_morra_cavalieri_hanifi_henningson_2020}.

Once the suitable modes are chosen, the energy scale needs to be applied. Considering the von Karman spectrum for isotropic homogeneous turbulence and following the 3D spectrum construction in \cite{Tennekes1972}, we have the formula for turbulent energy as a function of wave-number
\begin{equation}\label{eq:von_karman}
	E(\kappa) = \frac{2}{3} \frac{a (\kappa L)^4}{\left(b + (\kappa L)^2 \right)^{\frac{17}{6}}} L \cdot Tu^2, \quad L = \frac{1.8}{\kappa_{max}}
\end{equation}
where $a = 1.606$, $b = 1.35$ and $Tu$ is the turbulence intensity level defined as
\begin{equation}\label{eq:Tu}
	Tu = \sqrt{\frac{\left(u'^2_{rms}+v'^2_{rms}+w'^2_{rms}\right)}{3}}.
\end{equation}
In this equation, the integral length scale $L=7.5 \delta_0^*$ is set to \revTwo{the same value considered in \cite{sasaki_morra_cavalieri_hanifi_henningson_2020}, yielding a wavenumber of maximum energy, $\kappa_{max}$, near the minimum allowed value of $\kappa_l$. According to the results shown in \cite{brandt_schlatter_henningson_2004}, the increase of the turbulence integral length reduces the turbulence intensity decay at the free-stream and promotes transition in positions further upstream. Therefore, this choice of integral length scale consists of a worst case scenerio, which allows a shorter domain size in the streamwise direction.}

Concerning the energy scaling, it is demonstrated in \cite{Schlatter2001DirectNS} that the factor $\Phi$ in eq. (\ref{eq:expansion}) can be then expressed as
\begin{equation}\label{eq:energy_factor}
	\Phi(\kappa) = \sqrt{\frac{E(\kappa) \Delta \kappa}{N_s}},
\end{equation}
where $\Delta \kappa$ is the difference between consecutive values of $\kappa$.

Finally, the amplitudes of OSS modes in the continuous branch of the spectrum must be addressed at the top boundary of the domain. To prevent numerical instabilities, we multiply the eigenfunctions by a smooth step function $S(y)$ \cite{brandt_schlatter_henningson_2004} to dampen forcing perturbations above the position $y_d = 0.8 y_{max}$.

\revTwo{A more detailed discussion concerning the properties of the inflow perturbations generated using OSS modes in the continuous branch is presented in Appendix \ref{sec:propPert}.}

\section{Analysis techniques} \label{sec:techniques}

\subsection{Input-output formulation} \label{sec:in_out}
To apply the resolvent analysis framework over NS equations, we separate the velocity field in a two-dimensional, time-invariant, laminar solution \cite{jovanovic_bamieh_2005} or ensemble average flow \cite{mckeon_sharma_2010}
\begin{equation}
	\mathbf{U} = \left[U(x,y),V(x,y),0\right]^T
\end{equation}
and fluctuations
\begin{equation}
	\mathbf{u^\prime} = \left[u^\prime(x,y,z,t),v^\prime(x,y,z,t),w^\prime(x,y,z,t)\right]^T
\end{equation}
in order to write the linearised equations around $\mathbf{U}$, \revOne{as decribed in eq. \ref{eq:ns_1}}. Using tensor formulation, the system can be written as
\begin{equation}\label{eq:ns_2}
	\left.
	\begin{aligned}
		\frac{\partial u^\prime_i}{\partial t} + U_j \frac{\partial u^\prime_i}{\partial x_j} + u^\prime_j \frac{\partial U_i}{\partial x_j} = - \frac{\partial p^\prime}{\partial x_i} + \frac{1}{Re} \frac{\partial^2 u^\prime_i}{\partial x_j \partial x_j} + f_i + \sigma(\zeta_i - u^\prime_i),& \\
		\frac{\partial u^\prime_j}{\partial x_j} = 0,&
	\end{aligned}
	\right\}
\end{equation}
with non-linear terms grouped in $f_i = - u^\prime_j \frac{\partial u^\prime_i}{\partial x_j}$, considered in the resolvent framework as a forcing that drives the linear dynamics. The term $\zeta_i$ is the forcing vector defined in eq. (\ref{eq:ns_1}), which guarantees that inlet conditions in the model statistically match those observed in the boundary layer simulations. The function $\sigma(x)$ is the same \revOne{as} presented in figure \ref{fig:gain_fringe}.

Next, we apply the normal mode ansatz $\mathbf{u^\prime} = \mathbf{\hat{u}}(x,y) e^{i \left(\beta z - \omega t\right)}$ over velocity, pressure and forcing fields to expand eq. (\ref{eq:ns_2}) as
\begin{equation}\label{eq:lin_base}
	\left.
	\begin{aligned}
		-i \omega \hat{u} + U \frac{\partial \hat{u}}{\partial x} + V \frac{\partial \hat{u}}{\partial y} + \hat{u} \frac{\partial U}{\partial x} + \hat{v} \frac{\partial U}{\partial y} + \frac{\partial \hat{p}}{\partial x} - \frac{1}{Re}\left(\frac{\partial^2}{\partial x^2} -\beta^2 + \frac{\partial^2}{\partial y^2} \right) \hat{u} = \hat{f}_x + \sigma \left(\hat{\zeta}_x-\hat{u}\right),& \\
		-i \omega \hat{v} + U \frac{\partial \hat{v}}{\partial x} + V \frac{\partial \hat{v}}{\partial y} + \hat{u} \frac{\partial V}{\partial x} + \hat{v} \frac{\partial V}{\partial y} + \frac{\partial \hat{p}}{\partial y} - \frac{1}{Re}\left(\frac{\partial^2}{\partial x^2} -\beta^2 + \frac{\partial^2}{\partial y^2} \right) \hat{v} = \hat{f}_y  + \sigma \left(\hat{\zeta}_y-\hat{v}\right),& \\
		-i \omega \hat{w} + U \frac{\partial \hat{w}}{\partial x} + V \frac{\partial \hat{w}}{\partial y} + i \beta \hat{p} + \frac{1}{Re} \left(\frac{\partial^2}{\partial x^2} -\beta^2 + \frac{\partial^2}{\partial y^2} \right) \hat{w} = \hat{f}_z  + \sigma \left(\hat{\zeta}_z-\hat{w}\right),& \\
		\frac{\partial \hat{u}}{\partial x} + \frac{\partial \hat{v}}{\partial y} + i \beta \hat{w} = 0,&
	\end{aligned}
	\right\}
\end{equation}
where the non-linear term $\hat{f_i}$ can be written as a convolution of the Fourier transform of velocity components
\begin{equation}\label{eq:conv_f}
	\hat{f_i}(\beta,\omega) = - \hat{u}_j \ast \frac{\partial \hat{u}_i}{\partial x_j} = - \int_{-\infty}^{\infty} \int_{-\infty}^{\infty} \hat{u}_j\left(\beta_0,\omega_0\right) \: \frac{\partial }{\partial x_j} \hat{u}_i\left(\beta-\beta_0,\omega-\omega_0\right) \: d\beta_0 \: d\omega_0.
\end{equation}
This implies that $\hat{f_i}$ are the only terms responsible for energy transfers between different wavenumbers ($\beta,\beta_0,\beta-\beta_0$) and frequencies ($\omega,\omega_0,\omega-\omega_0$), in triads related to the turbulent energy cascade \cite{moffatt_2014,cheung_zaki_2014}.

\begin{figure}
	\centering
	\includegraphics[width=0.9\linewidth]{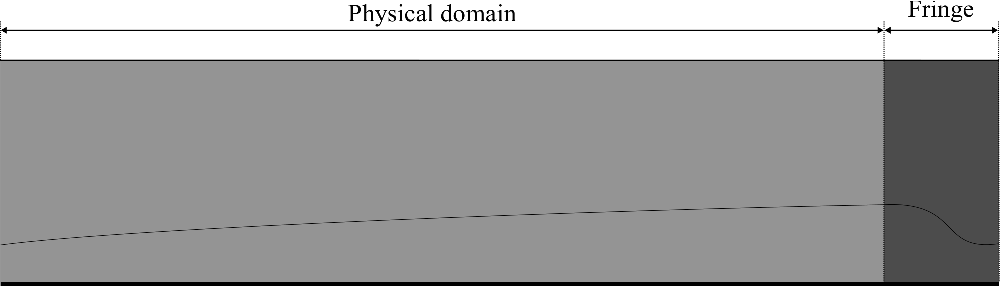}
	\caption{Diagram of the geometric distribution of input terms. While the non-linear term acts everywhere, the linear term is only present inside the fringe region. Legend: (Light gray) $\mathbf{B_f}\mathbf{\hat{f}}$; (Gray) $\mathbf{B_u} \mathbf{\hat{u}_r} + \mathbf{B_f}\mathbf{\hat{f}}$.}
	\label{fig:sketchbl_input}
\end{figure}

In practice, the fringe perturbation vector in Fourier space, $\mathbf{\hat{\zeta}}$, is approximated by the velocity fluctuations field computed from the simulations, denoted as $\mathbf{\hat{u}_r}$, which is substituted in equation \ref{eq:lin_base} for all three spatial components. Next, this equation is discretised reproducing the same grid of the LES and equivalent boundary conditions to write the system in state-space form
\begin{equation} \label{eq:state_space}
	\left.
	\begin{array}{r}
		\left(\mathbf{\Omega}+\mathbf{L}\right) \mathbf{\hat{q}} = \mathbf{B_u} \mathbf{\hat{u}_r} + \mathbf{B_f} \mathbf{\hat{f}}, \\
		\mathbf{\hat{y}} = \mathbf{H} \mathbf{\hat{q}},
	\end{array}
	\right\}
\end{equation}
and obtain
\begin{equation} \label{eq:resp}
	\mathbf{R} = \mathbf{H}\left(\mathbf{\Omega}+\mathbf{L}\right)^{-1} \implies \mathbf{\hat{y}} = \mathbf{R} \left(\mathbf{B_u} \mathbf{\hat{u}_r} + \mathbf{B_f} \mathbf{\hat{f}}\right),
\end{equation}
where $\mathbf{R}$ is the resolvent operator and $\mathbf{\hat{q}} = [\hat{u},\hat{v},\hat{w},\hat{p}]^T$, $\mathbf{\hat{y}} = [\hat{u},\hat{v},\hat{w}]^T$, $\mathbf{\hat{u}_r} = [\hat{u}_r,\hat{v}_r,\hat{w}_r]^T$, $\mathbf{\hat{f}} = [\hat{f}_x,\hat{f}_y,\hat{f}_z]^T$ are vectors composed of row-wise stacked components. Operators $\mathbf{\Omega}$, $\mathbf{L}$ $\mathbf{B_u}$, $\mathbf{B_f}$, $\mathbf{H}$ and boundary conditions are defined in appendix \ref{sec:linOp}. The operator $\mathbf{H}$ simply removes $\hat{p}$ from the output while the operator $\mathbf{B_u}$ restricts the application of the respective input to the region displayed in figure \ref{fig:sketchbl_input}. \revThree{It should be noted that the inclusion of the pressure, $p$, in the state $\mathbf{\hat{q}}$ removes the need to explicitly project the non-linear forcing, $\mathbf{\hat{f}}$, into a solenoidal space since the incompressible linearised Navier-Stokes system will redirect any non-solenoidal component in the non-linear forcing to the pressure field, as described in \cite{Rosenberg_2019}.} This formulation allows for the separation of contributions from external forcing, $\mathbf{\hat{y}_{L}}$, and non-linear forcing, $\mathbf{\hat{y}_N}$, as
\begin{equation}
	\left.
	\begin{array}{l}
		\mathbf{\hat{y}_{L}} = \mathbf{R}\mathbf{B_u} \mathbf{\hat{u}_r}, \\
		\mathbf{\hat{y}_N} = \mathbf{R} \mathbf{B_f} \mathbf{\hat{f}},
	\end{array}
	\right\}
\end{equation}
while still considering a single resolvent operator, \revOne{such that $\mathbf{\hat{y}} = \mathbf{\hat{y}_{L}} + \mathbf{\hat{y}_{N}} \approx \mathbf{\hat{u}_r}$. Even though the full response, $\mathbf{\hat{y}}$, is a superposition of linear and non-linear components, it is not the case that $\mathbf{\hat{y}_{L}}$ and $\mathbf{\hat{y}_{N}}$ evolve in a dynamically independent way, since $\mathbf{\hat{f}}$ is a function of the field fluctuations and needs to be computed before-hand from Navier-Stokes simulations in the context of the resolvent framework.} The component $\mathbf{B_u} \mathbf{\hat{u}_r}$ (linear input), accounts for the external flow perturbations coming through the domain upstream boundary and acts only in the fringe region, within a given pair $(\beta,\omega)$. On the other hand, $\mathbf{B_f} \mathbf{\hat{f}}$ (non-linear input), acts everywhere and accounts for the energy transfers between different wavenumbers and frequencies, due to the convolutional nature $\mathbf{\hat{f}}$, as described by eq. (\ref{eq:conv_f}).

\subsection{Spectral estimation}

Both $\mathbf{\hat{u}_r}$ and $\mathbf{\hat{f}}$ are computed directly from velocity fluctuations $\mathbf{u^\prime_r}$ from the simulation. Given the velocity fluctuations field $\mathbf{u^\prime}(x,y,z,t)$ at each snapshot, we compute non-linear terms $\mathbf{f}(x,y,z,t) = -\left(\mathbf{u^\prime_r} \cdot \nabla \right) \mathbf{u^\prime_r}$. Next, we apply the fast Fourier transform (FFT) in the periodic direction, $z$, to obtain $\mathbf{\bar{u}_r}(x,y,\beta,t)$ and $\bar{\mathbf{f}}(x,y,\beta,t)$.
These are organized in data matrices
\begin{equation} \label{eq:data_matrices}
	\mathbf{\bar{U}_r} = \left[
	\begin{array}{cccc}
		\mid & \mid & & \mid \\
		\mathbf{\bar{u}_r^{(1)}} & \mathbf{\bar{u}_r^{(2)}} & \cdots & \mathbf{\bar{u}_r^{(N_t)}} \\
		\mid & \mid & & \mid
	\end{array}\right], \quad
	\mathbf{\bar{F}} = \left[
	\begin{array}{cccc}
		\mid & \mid & & \mid \\
		\mathbf{\bar{f}^{(1)}} & \mathbf{\bar{f}^{(2)}} & \cdots & \mathbf{\bar{f}^{(N_t)}} \\
		\mid & \mid & & \mid
	\end{array}\right],
\end{equation}
each containing $N_t$ time-ordered snapshot column vectors.
The spectral estimation in frequency is performed using the Welch method \cite{1161901} via the algorithm presented in \cite{towne_schmidt_colonius_2018}. This procedure returns the quantities $\mathbf{\hat{u}_r}(x,y,\beta,\omega)$ and $\hat{\mathbf{f}}(x,y,\beta,\omega)$, which are assembled in the final spectral data matrices
\begin{equation} \label{eq:spectral_data_matrices}
	\mathbf{\hat{U}_r} = \left[
	\begin{array}{cccc}
		\mid & \mid & & \mid \\
		\mathbf{\hat{u}_r^{(1)}} & \mathbf{\hat{u}_r^{(2)}} & \cdots & \mathbf{\hat{u}_r^{(N_b)}} \\
		\mid & \mid & & \mid
	\end{array}\right], \quad
	\mathbf{\hat{F}} = \left[
	\begin{array}{cccc}
		\mid & \mid & & \mid \\
		\mathbf{\hat{f}^{(1)}} & \mathbf{\hat{f}^{(2)}} & \cdots & \mathbf{\hat{f}^{(N_b)}} \\
		\mid & \mid & & \mid
	\end{array}\right],
\end{equation}
for each pair wavenumber and frequency $(\beta,\omega)$, containing $N_b$ columns which correspond to the number of blocks used in the windowing procedure.

\subsection{Spectral correction due to windowing} \label{sec:spectral_correction}

The presence of the windowing function in the spectral estimation adds new terms to the response of the LNS equations written in eq. (\ref{eq:ns_2}), as pointed out by \cite{martini2020accurate}. Considering the operators defined in Appendix \ref{sec:linOp} and the matrices in eq. (\ref{eq:data_matrices}), we write eq. (\ref{eq:state_space}) in the time domain as
\begin{equation}\label{eq:state_space_time}
	\left.
	\begin{array}{r}
		\mathbf{B_y} \frac{\partial \mathbf{y}}{\partial t} + \mathbf{L} \mathbf{q} = \mathbf{B_u} \mathbf{\bar{u}_r} + \mathbf{B_f} \mathbf{\bar{f}}, \\
		\mathbf{y} = \mathbf{H} \mathbf{q},
	\end{array}
	\right\} , \quad \mathbf{B_y} = \mathbf{B_f}.
\end{equation}

Applying the Welch method for spectral estimation implies that each data block is multiplied by a windowing function $w(t)$ so that eq. (\ref{eq:state_space_time}) becomes
\begin{equation}\label{eq:ss_windowed}
	\left.
	\begin{array}{r}
		w \mathbf{B_y} \frac{\partial \mathbf{y}}{\partial t} + w \mathbf{L} \mathbf{q} = w \mathbf{B_u} \mathbf{\bar{u}_r} + w \mathbf{B_f} \mathbf{\bar{f}} \\
		w \mathbf{y} = w \mathbf{H} \mathbf{q}
	\end{array}
	\right\}.
\end{equation}
The windowing function $w(t)$ commutes with all time-invariant operators, for instance,
\begin{equation}\label{eq:prop1}
	w \mathbf{L} \mathbf{q} = \mathbf{L} \left(w \mathbf{q}\right),
\end{equation}
but not with the time derivative, which obeys the identity
\begin{equation}\label{eq:deriv_window}
	w \frac{\partial \mathbf{y}}{\partial t} = \frac{\partial}{\partial t} \left(w \mathbf{y}\right) - \frac{d w}{d t} \mathbf{y}.
\end{equation}

These relations imply that eq. (\ref{eq:ss_windowed}) can be rewritten in the form
\begin{equation}
	\left.
	\begin{array}{r}
		\mathbf{B_y} \frac{\partial}{\partial t} \left(w \mathbf{y}\right) + \mathbf{L} \left(w \mathbf{q}\right) = \mathbf{B_u} \left(w \mathbf{\bar{u}_r}\right) + \mathbf{B_f} \left(w \mathbf{\bar{f}}\right) + \mathbf{B_y} \left(\frac{d w }{d t} \mathbf{y}\right),\\
		w\mathbf{y} = \mathbf{H} \left(w \mathbf{q}\right),
	\end{array}
	\right\}
\end{equation}
and transformed into frequency space, as
\begin{equation}
	\left.
	\begin{array}{r}
		\left(\mathbf{\Omega}+\mathbf{L}\right) \mathbf{\hat{q}} = \mathbf{B_u} \mathbf{\hat{u}_r} + \mathbf{B_f} \mathbf{\hat{f}} + \mathbf{\hat{q}_c}, \\
		\mathbf{\hat{y}} = \mathbf{H} \mathbf{\hat{q}},
	\end{array}
	\right\}
\end{equation}
which contains a windowing correction term
\begin{equation}
	\mathbf{\hat{q}_c} = \mathbf{B_y} \: \mathcal{F} \left\{\frac{d w }{d t} \mathbf{y}\right\},
\end{equation}
where $\mathcal{F}$ denotes the Fourier transform in time. In practice, $\mathbf{\hat{q}_c}$ is constructed using available simulation data
\begin{equation}
	\mathbf{\hat{q}_c} \equiv \mathbf{B_y} \: \mathcal{F} \left\{\frac{d w }{d t} \mathbf{\bar{u}_r}\right\},
\end{equation}
with $\mathbf{\bar{u}_r}$ representing the column vectors of $\mathbf{\bar{U}_r}$, in eq. (\ref{eq:data_matrices}). The term $\frac{d w }{d t}$ is computed directly from the analytical formula of the windowing function used in the Welch method.

Physically, $\mathbf{\hat{q}_c}$ is related to transients that are inevitably introduced when the signal is windowed (i.e., inputs necessary to match initial and final conditions of each data block) and implies that windowed spectral estimations create a mismatch between inputs and outputs, even in the case of perfectly converged statistics. Even though the windowing procedure cannot be avoided when dealing with large datasets, due to computer memory constraints, the magnitude of the correction term $\mathbf{\hat{q}_c}$ can be reduced by increasing the size of the data block, which tends to proportionally decrease the value of $d w /d t$ since longer blocks imply wider windows with smaller derivatives.

\subsection{Response reconstruction from inputs} \label{sec:rec_inputs}

From eq. (\ref{eq:resp}) and the spectral data matrices in eq. (\ref{eq:spectral_data_matrices}), we can compute the reconstructed response in Fourier space
\begin{equation} \label{eq:total_output}
	\mathbf{\hat{Y}}= \mathbf{\hat{Y}_L} + \mathbf{\hat{Y}_N} + \mathbf{\hat{Y}_C} = \mathbf{R}\mathbf{B_u} \mathbf{\hat{U}_r} + \mathbf{R}\mathbf{B_f} \mathbf{\hat{F}} + \mathbf{R} \mathbf{\hat{Q}_c},
\end{equation}
where $\mathbf{\hat{Q}_C}$ is the correction due to windowing, discussed in section \ref{sec:spectral_correction}, in order to obtain $\mathbf{\hat{Y}} \approx \mathbf{\hat{U}_r}$ by construction. In other words, the sum of all inputs with the proper correction of the distortions generated by the windowing procedure leads, in principle, to the recovery of the simulated velocity fluctuation fields, by means of the resolvent operator. This allows us to calculate separate contributions of linear mechanisms resulting from the upstream fluctuations, related to $\mathbf{\hat{U}_r}$, and non-linear receptivity due to triadic interactions, related to $\mathbf{\hat{F}}$.

The cross-spectral density (CSD) matrix of $\mathbf{\hat{Y}}$, can be estimated from the ensemble as
\begin{equation}\label{eq:CSD}
	\mathbf{\hat{C}_{YY}} = \frac{1}{N_b} \mathbf{\hat{Y}}\mathbf{\hat{Y}}^H = \frac{1}{N_b} \mathbf{\hat{Y}}\left(\mathbf{\hat{Y}_L} + \mathbf{\hat{Y}_N} + \mathbf{\hat{Y}_C}\right)^H,
\end{equation}
with the superscript $\{\cdot\}^H$ representing the conjugate transpose, and can be rewritten as
\revOne{\begin{equation} \label{eq:CSD_factors}
		\left. \begin{array}{ccccccc}
			\mathbf{\hat{C}_{YY_L}} & = & \mathbf{\hat{C}_{Y_L Y_L}} & + & \mathbf{\hat{C}_{Y_{NL} Y_L}} & + & \mathbf{\hat{C}_{Y_C Y_L}} \\
			\mathbf{\hat{C}_{YY_{NL}}} & = & \mathbf{\hat{C}_{Y_L Y_{NL}}} & + & \mathbf{\hat{C}_{Y_{NL} Y_{NL}}} & + & \mathbf{\hat{C}_{Y_C Y_{NL}}}\\
			\mathbf{\hat{C}_{YY_C}} & = & \mathbf{\hat{C}_{Y_L Y_C}} & + & \mathbf{\hat{C}_{Y_{NL} Y_C}} & + & \mathbf{\hat{C}_{Y_C Y_C}}
		\end{array} \right\},
\end{equation}}
\begin{equation}\label{eq:CSD_factors2}
	\mathbf{\hat{C}_{YY}} = \mathbf{\hat{C}_{YY_L}} + \mathbf{\hat{C}_{YY_N}} + \mathbf{\hat{C}_{YY_C}}.
\end{equation}

Each one of the three factors in \eqref{eq:CSD_factors2} computes the coherence between the respective response component and the reconstructed signal. Even though these are not independent quantities, since factors contain cross-products between components, this formulation constitutes a budget measure of how each component contributes to the spectrum of the reconstructed signal.

In practice, the CSD matrix, $\mathbf{\hat{C}_{YY}}$, is never fully assembled due to its huge size. Since we are interested in the kinetic energies at each pair $(\beta,\omega)$, we only effectively compute the power spectral density (PSD), defined as the diagonal of the CSD matrix. Considering that the PSD is always positive and real, we obtain the relations
\begin{equation}\label{eq:stat_psd}
	\mathbf{P_U} = Re\left\{ \text{diag} \left(\mathbf{\hat{C}_{U_r U_r}}\right) \right\},
\end{equation}
\begin{equation} \label{eq:budget}
	\begin{aligned}
		\mathbf{P_Y}&= Re\left\{ \text{diag} \left(\mathbf{\hat{C}_{YY_L}}\right) \right\} + Re\left\{ \text{diag} \left(\mathbf{\hat{C}_{YY_N}}\right) \right\} + Re\left\{ \text{diag} \left(\mathbf{\hat{C}_{YY_C}}\right) \right\} \\
		& = \mathbf{\Pi_L} + \mathbf{\Pi_N} + \mathbf{\Pi_C}
	\end{aligned},
\end{equation}
where $\mathbf{P_U} \approx \mathbf{P_Y}$ by construction. The term $\mathbf{P_U}$, computed directly from the velocity fluctuation fields of the simulation, data matrix $\mathbf{\hat{U}_r}$ in eq. (\ref{eq:spectral_data_matrices}), is called statistical PSD. Then, $\mathbf{P_Y}$, computed through the sum of components of the input-output model is called reconstructed PSD. Because of the cross products, $\mathbf{\Pi}$ components are not PSDs and can assume either positive or negative values, which are interpreted, respectively, as inflows or outflows of energy at a given pair $(\beta,\omega)$, \revOne{i.e.,
energy exchanges between linear, non-linear and correction components.}

The equivalence between $\mathbf{P_U}$ and $\mathbf{P_Y}$ is verified numerically by the reconstruction coefficient defined as
\begin{equation}\label{eq:coeff_PSD}
	\gamma(\beta,\omega) = \frac{\mathbf{P_U}^T \mathbf{P_Y}}{\mathbf{P_U}^T \mathbf{P_U}}.
\end{equation}
Within this metric, a coefficient $\gamma \approx 1$ indicates that the reconstruction $\mathbf{P_Y}$ has the correct magnitude and shape, implying that the input-output model is accurate. Thus, linear and non-linear components, $\mathbf{\Pi_L}$ and $\mathbf{\Pi_N}$ respectively, are representative in the system's response, \textit{assuming they are individually more significant than the windowing correction term, $\mathbf{\Pi_C}$}. We may thus assess, using simulation data and resolvent analysis, the relative contribution of linear and non-linear mechanisms in disturbance growth.

To reduce the quantity of data presented, only $\mathbf{\Pi_L}$ and $\mathbf{\Pi_N}$ components of $\mathbf{P_Y}$ will be displayed in corresponding the results. Proof that conditions exposed in the previous paragraph are met is given by presenting the associated coefficient $\gamma$ and the magnitude of the correction component, defined as $\max\left|\mathbf{\Pi_C}\right|$, for each spatial direction. A more complete comparison between statistical and reconstructed PSDs for selected pairs $(\beta,\omega)$ is exposed in Appendix \ref{sec:compPSD}.

\subsection{Resolvent-based extended spectral POD} \label{sec:respectral POD}

The resolvent-based extended spectral POD (RESPOD) presented in \cite{karban_martini_cavalieri_lesshafft_jordan_2022} is a form of extended POD \cite{Boree2003} which exploits the dynamical properties of spectral POD \cite{towne_schmidt_colonius_2018} to statistically correlate inputs and outputs of a linear system in frequency space. \revThree{The method can be viewed as a procedure to obtain forcing modes, ranked by their effect on the most energetic flow structures. These can be employed, for instance, in turbulence control models, as in \cite{chevalier_hœpffner_bewley_henningson_2006}.}

Given input and output spectral data matrices, respectively $\mathbf{\hat{F}}$ and $\mathbf{\hat{U}}$, related linearly in the resolvent framework by
\begin{equation} \label{eq:resolvent_Ur}
	\mathbf{\hat{U}} = \mathcal{R} \mathbf{\hat{F}},
\end{equation}
we define an augmented state
\begin{equation}\label{eq:expanded_state}
	\mathbf{\hat{Q}} =
	\left[
	\begin{array}{c}
		\mathbf{\hat{U}} \\
		\mathbf{\hat{F}}
	\end{array}
	\right]
\end{equation}
over which we apply the spectral POD method using the snapshot algorithm \cite{sirovich1987}. By computing the weighted CSD matrix $\mathbf{\hat{M}_Q}$ in the row-space of $\mathbf{\hat{Q}}$, we have
\begin{equation}\label{eq:Mq}
	\mathbf{\hat{M}_Q} = \frac{1}{N_b} \mathbf{\hat{Q}}^H \left[ \begin{matrix} \mathbf{W} & 0 \\ 0 & 0 \end{matrix} \right] \mathbf{\hat{Q}} = \frac{1}{N_b} \mathbf{\hat{U}}^H \mathbf{W} \mathbf{\hat{U}},
\end{equation}
where matrix $\mathbf{W}$ represents the grid quadrature weights. Next, we arrive at the eigenproblem
\begin{equation}\label{eq:snapshot_eig}
	\mathbf{\hat{M}_Q} \mathbf{\Theta} = \mathbf{\Theta} \mathbf{\Lambda},
\end{equation}
where $\mathbf{\Theta}$ and $\mathbf{\Lambda}$ are, respectively, spectral POD expansion coefficients and energies of $\mathbf{\hat{U}}$. The respective eigenvectors in the column-space of $\mathbf{\hat{Q}}$ are then given by
\begin{equation}\label{eq:snapshot_res}
	\mathbf{\tilde{\Psi}} = \left[ \begin{matrix} \mathbf{\Psi} \\ \mathbf{\Phi} \end{matrix} \right] = \frac{1}{\sqrt{N_b}} \mathbf{\hat{Q}} \mathbf{\Theta} \mathbf{\Lambda}^{-1/2},
\end{equation}
showing that the augmented eigenvector $\mathbf{\tilde{\Psi}}$ is composed of spectral POD modes $\mathbf{\Psi}$ and forcing modes $\mathbf{\Phi}$, which are both directly computed from the expansion coefficients $\mathbf{\Theta}$ and energies/eigenvalues $\mathbf{\Lambda}$. Finally, by substituting eqs. (\ref{eq:resolvent_Ur}) and (\ref{eq:expanded_state}) into eq. (\ref{eq:snapshot_res}), we get the RESPOD relation
\begin{equation}\label{eq:respectral POD}
	\mathbf{\Psi} = \mathcal{R} \mathbf{\Phi},
\end{equation}
which shows that response and forcing modes are related by the resolvent operator.

If $\mathcal{R}$ is non-singular, $\mathbf{\Phi}$ is simply the application of the inverse operator $\mathcal{R}^{-1}$ over $\mathbf{\Psi}$. However, if $\mathcal{R}$ is singular, $\mathbf{\Phi}$ can be shown to contain both minimal-norm forcing components, the same computed by resolvent-based estimation from \cite{towne_lozano-duran_yang_2020} and \cite{martini_cavalieri_jordan_towne_lesshafft_2020}, and dynamically unobservable components, which are correlated to the minimal-norm forcing in the subspace spanned by the input signal \cite{karban_martini_cavalieri_lesshafft_jordan_2022}. One thus obtains forcing modes $\mathbf{\Phi}$, taken from data, which drive the observed response spectral POD modes $\mathbf{\Psi}$. This yields a ranked modal decomposition of non-linear forcing data statistics, which is a particularly useful modelling tool in cases where the response results predominantly from the non-linear dynamics.

\subsection{Spectral parameters}

Spectral estimation via the Welch method is performed using blocks of $N_{FFT}=192$ realisations, a value defined via the cross-correlation procedure detailed in \cite{blanco_martini_sasaki_cavalieri_2022}. In all the following analyses, we employ a windowing function
\begin{equation}\label{eq:windowing_function}
	w(t) = \sin\left(\frac{\pi t}{T}\right), \quad t \in [0,T]
\end{equation}
and overlap of $O_{FFT} = 3/4$ between consecutive blocks, based on the guidance given in the work of \cite{5200747}.


\section{Statistical power spectrum} \label{sec:spectra}

\begin{figure}
	\centering
	\begin{subfigure}[b]{.5\textwidth}
		\centering
		\includegraphics[width=\linewidth]{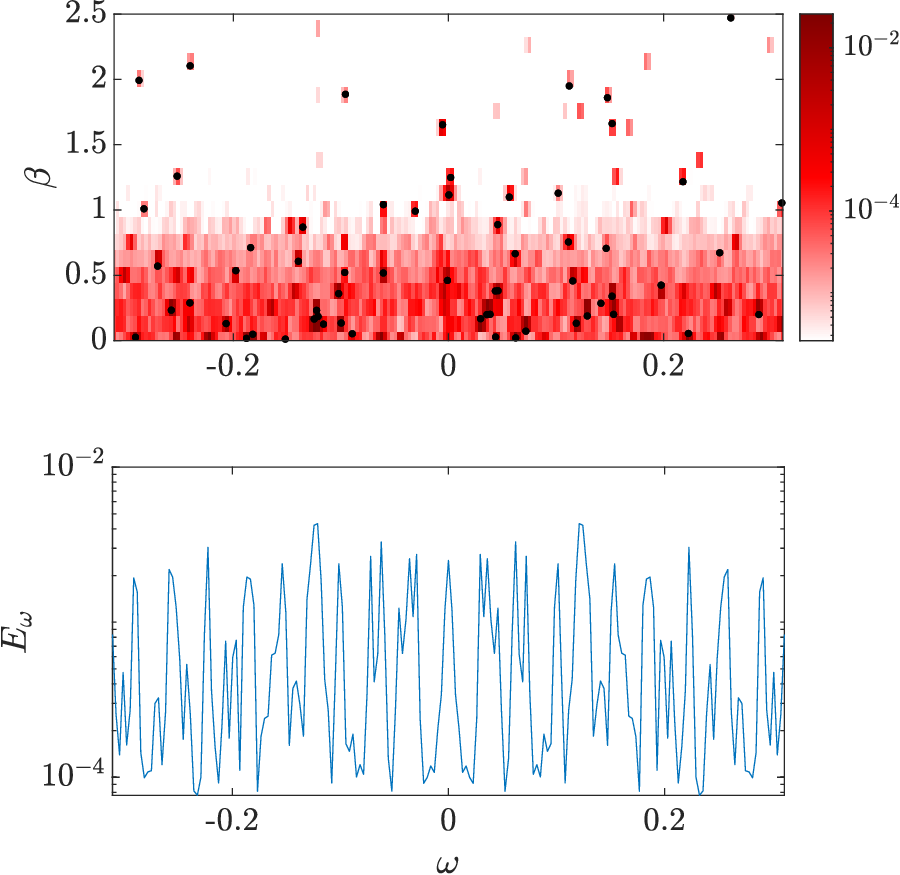}
		\caption{$Tu = 0.5\%$}
		\label{fig:specTu05}
	\end{subfigure}%
	\begin{subfigure}[b]{.5\textwidth}
		\centering
		\includegraphics[width=\linewidth]{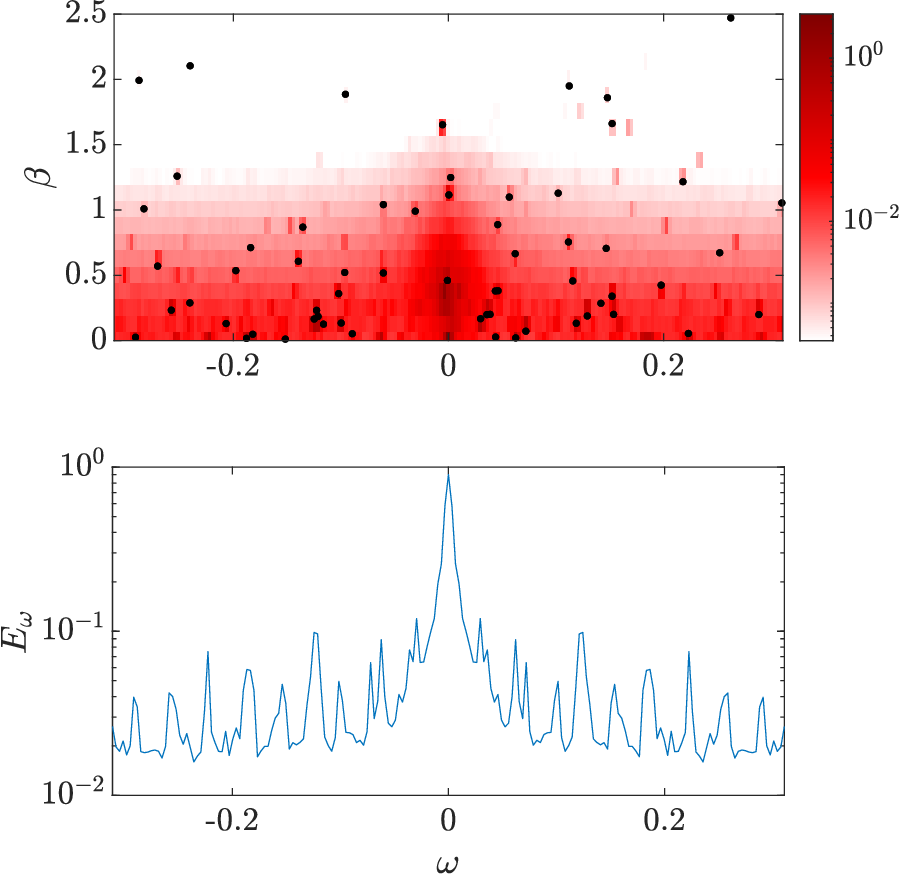}
		\caption{$Tu = 3.5\%$}
		\label{fig:specTu35}
	\end{subfigure}
	\caption{Energy spectra computed from snapshots. (Top) Energy $E(\beta,\omega)$ distributed over all wavenumbers and frequencies. Black dots indicate the spectrum of the FST applied at the fringe; (Bottom) Energy spectrum in frequency, with summation considering positive and negative wavenumbers $\beta$.}
	\label{fig:spectra}
\end{figure}

In the first analysis, we compute the statistical PSD, $\mathbf{P_U}$, at each pair $(\beta,\omega)$, for all the available FST levels, and subsequently integrate over all $N$ spatial points within the physical domain (excluding the fringe), in the $x$ and $y$ directions
\begin{equation}
	E(\beta,\omega) = \sum_{i = 1}^{N} \left(\mathbf{W} \mathbf{P_U}\right)_i,
\end{equation}
and over resolved wavenumbers
\begin{equation}
	E_\omega = \sum_\beta E(\beta,\omega) \: \Delta \beta.
\end{equation}
to compute the corresponding kinetic energy spectrum. Here, the matrix $\mathbf{W}$, which is also present in eq. (\ref{eq:Mq}), absorbs the terms $\Delta x$ and $\Delta y$ of the Riemann sum.  The resulting data, presented in figure \ref{fig:spectra}, clarifies that the amplification generated by the increase of FST levels is concentrated around the near-zero frequencies, as expected for streaks \cite{brandt_schlatter_henningson_2004}, following the optimal growth theory \cite{luchini2000}. However, one interesting observation is that the peak in the energy spectrum for the case of $Tu = 3.5\%$ does not coincide with the spectrum of the FST applied at the fringe. This is the first indication of the existence of non-linear mechanisms promoting the growth of perturbations.

\begin{figure}
	\centering
	\begin{subfigure}[b]{.5\textwidth}
		\centering
		\includegraphics[width=0.95\linewidth]{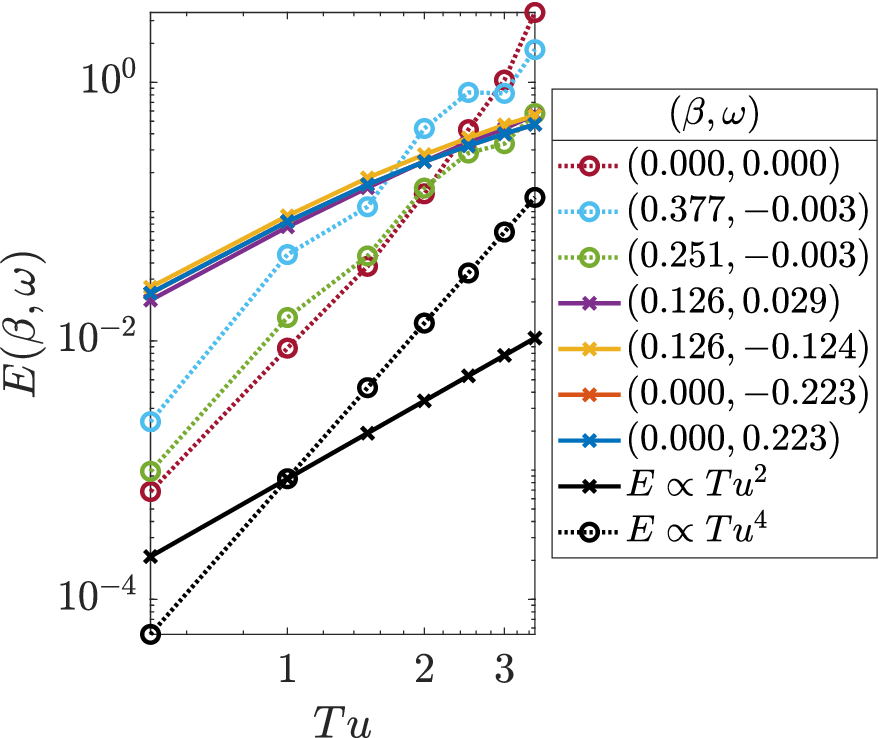}
		\caption{}
		\label{fig:peaks_Tu}
	\end{subfigure}%
	\begin{subfigure}[b]{.5\textwidth}
		\centering
		\includegraphics[width=0.95\linewidth]{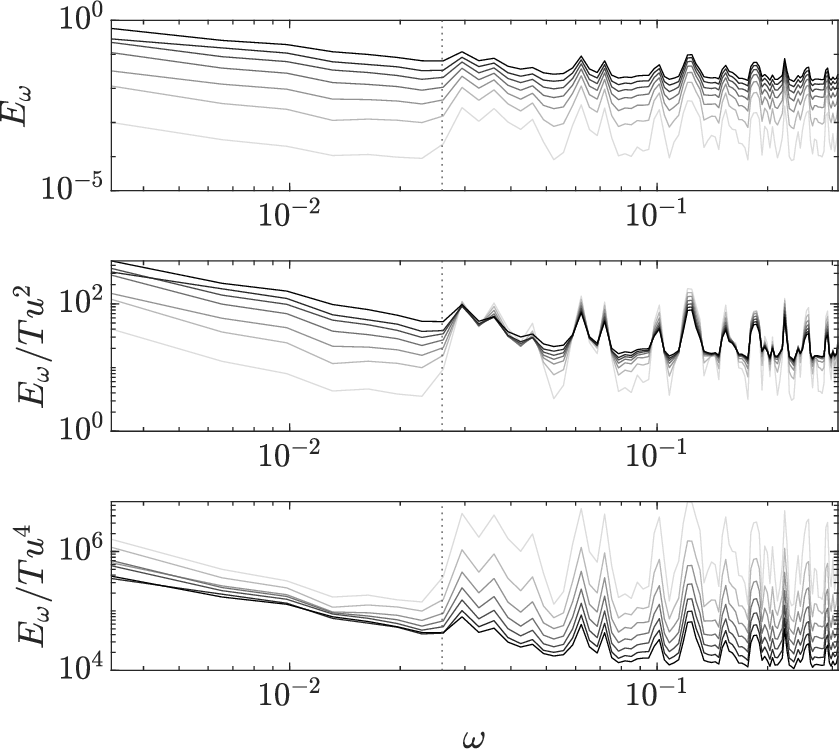}
		\caption{}
		\label{fig:spec_omega}
	\end{subfigure}
	\caption{Energy as a function of FST level, $Tu$. (a) Evolution of most energetic pairs $(\beta,\omega)$. Legend: ($\times$) Linear behaviour; ($\circ$) Non-linear behaviour. (b) Normalised energy spectrum. Higher frequencies grow with a $Tu^2$ dependency, while lower frequencies scale with a factor closer to $Tu^4$. Darker lines indicate higher FST levels. The vertical line, drawn at $\omega = 0.026$, separates higher and lower frequency ranges.}
	\label{fig:evo_Tu}
\end{figure}

Next, we sort the four most energetic pairs $(\beta,\omega)$ for each available $Tu$. By plotting the evolution of the identified pairs (figure \ref{fig:peaks_Tu}), we observe two distinct behaviours. For higher frequencies, energies grow at a rate closely proportional to $Tu^2$, implying linear dependency concerning the incoming turbulent energy. On the other hand, near-zero frequencies display a faster energy growth, suggesting non-linear dependence on the incoming FST.

These same conclusions can be drawn by normalising the energy $E_\omega$ by powers of $Tu$, as shown in figure \ref{fig:spec_omega}. We see that, indeed, higher frequencies, $|\omega| > 0.026$, collapse when normalised by $Tu^2$ while lower frequencies, $|\omega| < 0.026$, require larger exponents, closer to the expected $Tu^4$ resulting from the quadratic nature of the non-linear term. 

\section{Reconstructed power spectrum}

\begin{figure}
	\centering
	\includegraphics[width=\linewidth]{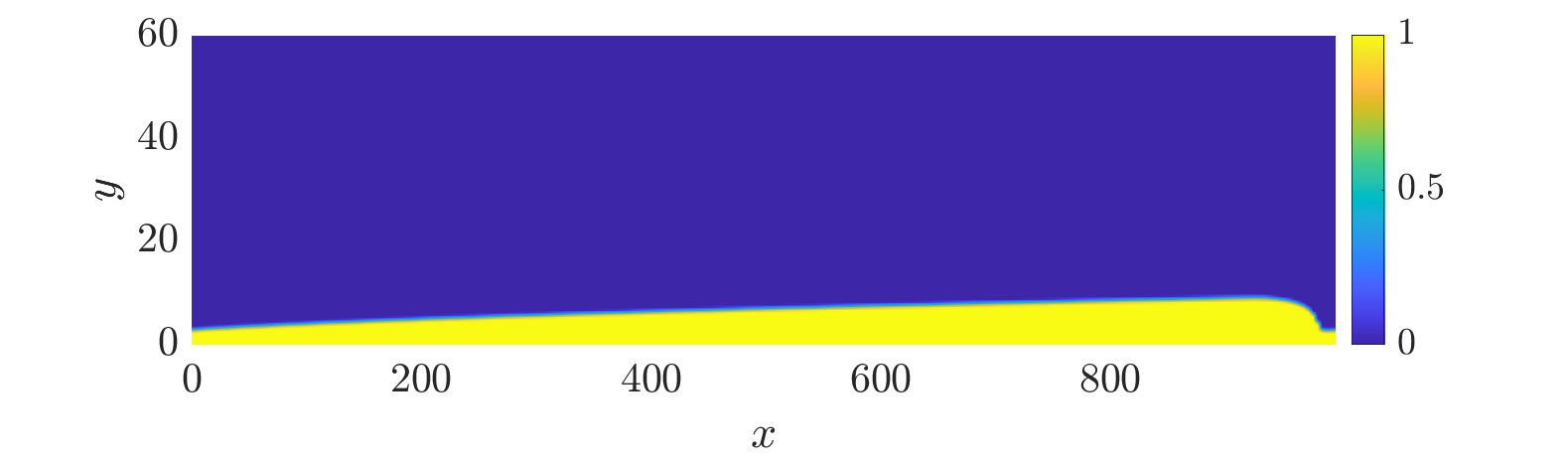}
	\caption{Spatial distribution of the boundary layer mask, $\mathbf{W_{BL}}$.}
	\label{fig:blmask}
\end{figure}

\begin{figure}
	\centering
	\includegraphics[width=\linewidth]{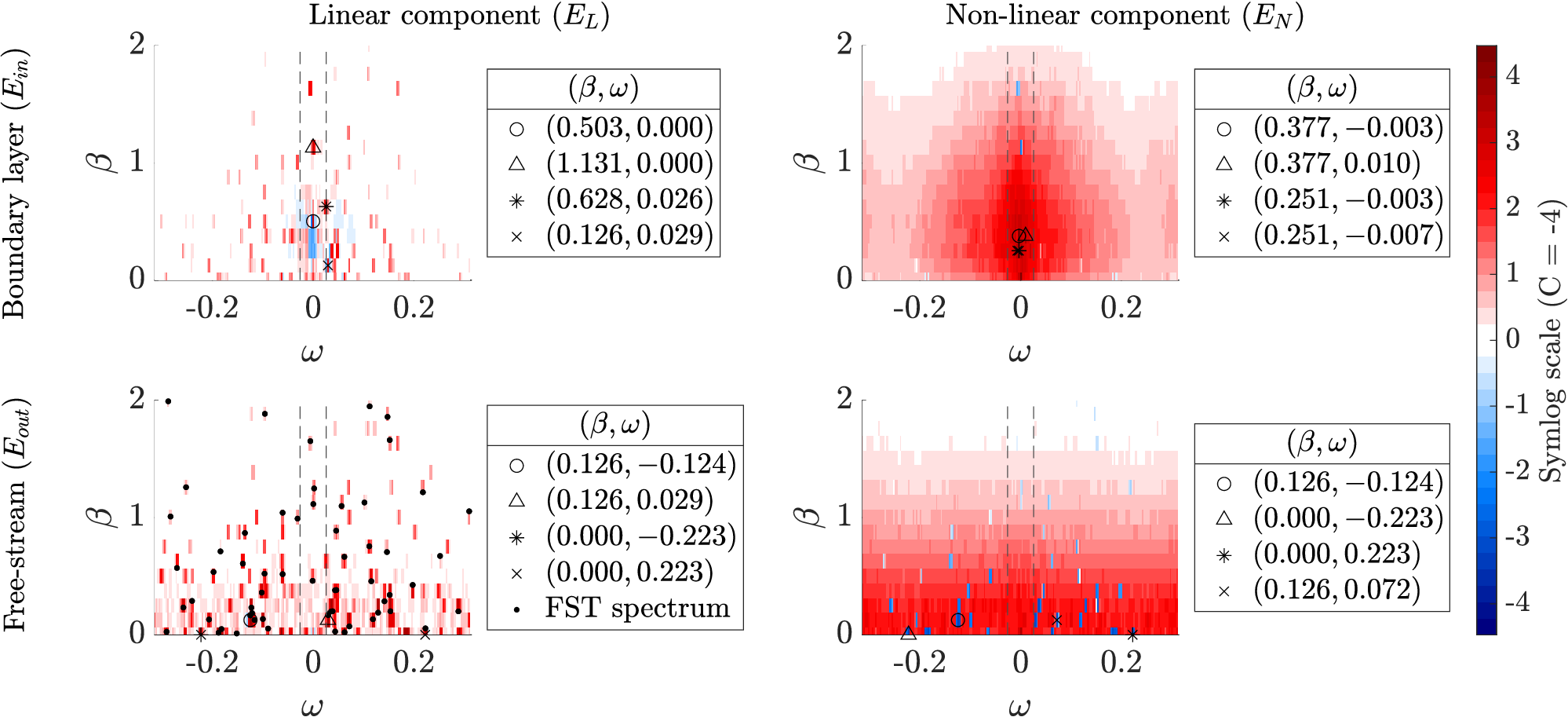}
	\caption{Energy spectra of the reconstructed PSD components, inside and outside the boundary layer. Colours in the symlog scale. Pairs $(\beta,\omega)$ are listed in descending order of energy magnitude, excluding the mean. Black dots indicate the incoming FST spectrum. Vertical lines drawn at $|\omega| = 0.026$ separate higher and lower frequency ranges.}
	\label{fig:spectraComp}
\end{figure}

In a subsequent investigation, we focus on the case $Tu=3.5\%$ and seek to understand which of the pairs is more related to linearly/non-linearly generated structures near the wall. For this, we compute the components $\mathbf{\Pi_L}$ and $\mathbf{\Pi_N}$ of the reconstructed PSD, which are then integrated into two separated regions in space, divided at the Blasius boundary layer $\delta_{99}$ thickness position. Therefore, we obtain inside and outside linear contribution components
\begin{equation}
	E_{L,in}(\beta,\omega) = \sum_{i = 1}^{N} \left(\mathbf{W_{BL}} \mathbf{W} \mathbf{\Pi_L}\right)_i,
\end{equation}
\begin{equation}
	E_{L,out}(\beta,\omega) = \sum_{i = 1}^{N} \left(\left(\mathbf{I}-\mathbf{W_{BL}}\right)\mathbf{W}\mathbf{\Pi_L}\right)_i,
\end{equation}
and, analogously, non-linear contribution components
\begin{equation}
	E_{N,in}(\beta,\omega) = \sum_{i = 1}^{N} \left(\mathbf{W_{BL}}\mathbf{W} \mathbf{\Pi_N}\right)_i,
\end{equation}
\begin{equation}
	E_{N,out}(\beta,\omega) = \sum_{i = 1}^{N} \left(\left(\mathbf{I}-\mathbf{W_{BL}}\right)\mathbf{W}\mathbf{\Pi_N}\right)_i,
\end{equation}
where the term $\mathbf{W_{BL}}$ is the boundary layer mask, a diagonal weight matrix constructed with the same ordering as $\mathbf{W}$, whose spatial distribution is displayed in figure \ref{fig:blmask}. The data resulting from this procedure is exposed in figure \ref{fig:spectraComp}. Since $\mathbf{\Pi}$ components can assume both positive and negative values, the spectrum is plotted in the symmetric log scale \cite{webber2012bi}.

\revThree{Outside the boundary layer, from the superposition with the introduced FST spectrum (OSS modes), we perceive that the $E_{L,out}$ spectrum is heavily influenced by the perturbations introduced in the fringe zone. Overall, the energy distribution in $E_{L,out}$ is discrete and reflects the finite set of modes superposed to create the incoming FST. Every OSS mode matches with an energy peak, even though peaks without a corresponding OSS mode exist (see Appendix \ref{sec:propPert}). Indeed, the most energetic peaks in the higher frequency range generally fall over the FST spectrum.}

Moreover, it is noteworthy that peaks in the $E_{L,out}$ spectrum often coincide with negative energy contributions in the $E_{N,out}$ spectrum. This implies that the turbulent energy introduced to a given wavenumber-frequency combination via linear mechanisms is transfered to other wavenumbers and frequencies via triadic interactions, in a mechanism characteristic of the turbulent energy cascade. This non-linear mechanism also explains the broad distribution in frequency and wavenumber in the $E_N$ spectrum contrasting with the more discrete peaks featured in the $E_L$ spectrum.

\revThree{Inside the boundary layer, the most energetic pairs are identified within the lower frequency range, in both linear and non-linear spectra. This feature is consistent with the optimal growth theory devised in \cite{doi:10.1063/1.869908} \revOne{and} \cite{luchini2000}, which states that boundary layer disturbances are optimally amplified for nearly zero frequencies by the linearised NS operator. On the other hand, the boundary layer acts as a barrier to the penetration of rapidly changing perturbations \cite{10.1063/1.869716}, an effect that is noticeable in the spectrum through the lower energy content in the high-frequency range of $E_{in}$ when compared to $E_{out}$.}

\revThree{Since the linear dynamics are decoupled in wavenumber and frequency, the energy peaks inside of $E_{L,in}$ must match the peaks in $E_{L,out}$. If the energy peak in $E_{L,in}$ is the predominant component of the energy inside the boundary layer, we conclude that near-wall structures were induced by linear interactions with the incoming FST and, therefore, are subject to linear receptivity mechanisms. However, if the energy peak in $E_{N,in}$ is predominant, the energy to excite near-wall structures inevitably comes from the interaction with other wavenumbers and frequencies through the non-linear forcing term, and thus there exists a non-linear receptivity mechanism. For the case presented, $E_{N,in}$ spectrum has a strong peak at $(\beta,\omega) = (0.377,-0.003)$, which contains an order of magnitude more energy than all surrounding pairs, while the $E_{L,in}$ has two distinct zero-frequency peaks at $(0.503,0.000)$ and $(1.131,0.000)$.}

The next sections in this work will analyse specific wavenumbers and frequencies, found to be relevant for the transition dynamics. Using the energy criteria, we focus on the pairs $(\beta,\omega) = (0.126,-0.124)$, which is most important in FST levels below $2.0\%$, and $(0.377,-0.003)$, most important from $Tu = 2.0 \%$ and above, excluding the mean. To this list, we add the zero-frequency pairs $(0.503,0.000)$ and $(1.131,0.000)$, identified in the spectrum of $E_{L,in}$, whose roles will be further discussed.

\section{Free stream structures} \label{sec:freestream}

\begin{figure}
	\centering
	\begin{subfigure}[b]{.49\textwidth}
		\centering
		\includegraphics[width=\linewidth]{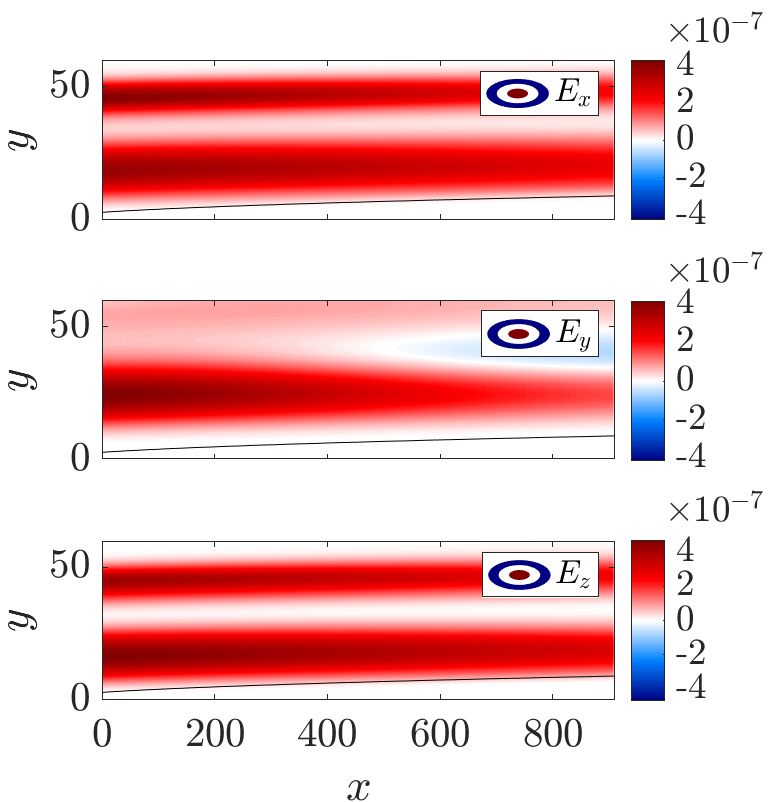}
		\caption{Linear component, $\mathbf{\Pi_{L}}$}
		\label{fig:Clin_Tu05_omega-0.1244_beta0.1257}
	\end{subfigure}%
	\begin{subfigure}[b]{.49\textwidth}
		\centering
		\includegraphics[width=\linewidth]{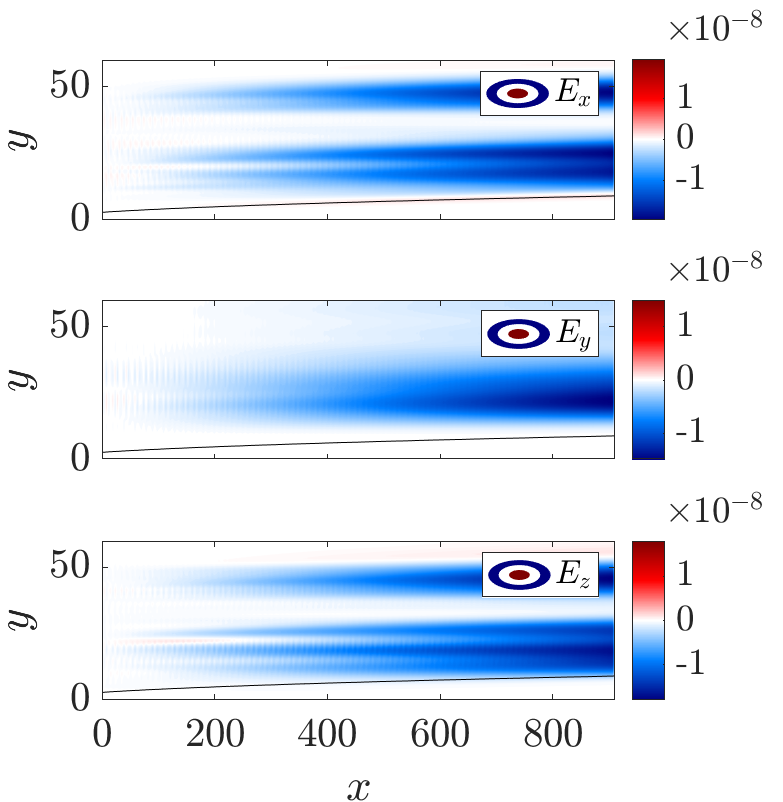}
		\caption{Non-linear component, $\mathbf{\Pi_N}$}
		\label{fig:Cnlin_Tu05_omega-0.1244_beta0.1257}
	\end{subfigure}
	\caption{Components of $\mathbf{P_Y}$, for $(\beta,\omega)=(0.126,-0.124)$ and $Tu=0.5\%$. The linear response is dominant while the non-linear one is negligible. Parameters: $\gamma = 0.951$; $\max\left(\left|\mathbf{\Pi_C}\right|\right) = \left[7.2 \times 10^{-8},2.0 \times 10^{-7},5.3 \times 10^{-8}\right]$.}
	\label{fig:Comp_Tu05_omega-0.1244_beta0.1257}
\end{figure}

\begin{figure}
	\centering
	\begin{subfigure}[b]{.49\textwidth}
		\centering
		\includegraphics[width=\linewidth]{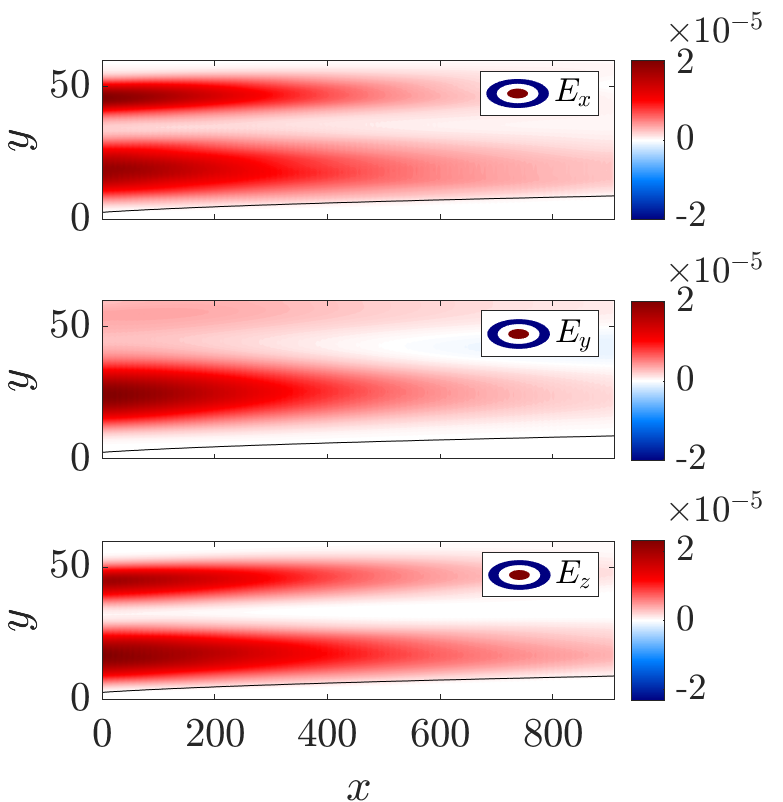}
		\caption{Linear component, $\mathbf{\Pi_L}$}
		\label{fig:Clin_Tu35_omega-0.1244_beta0.1257}
	\end{subfigure}%
	\begin{subfigure}[b]{.49\textwidth}
		\centering
		\includegraphics[width=\linewidth]{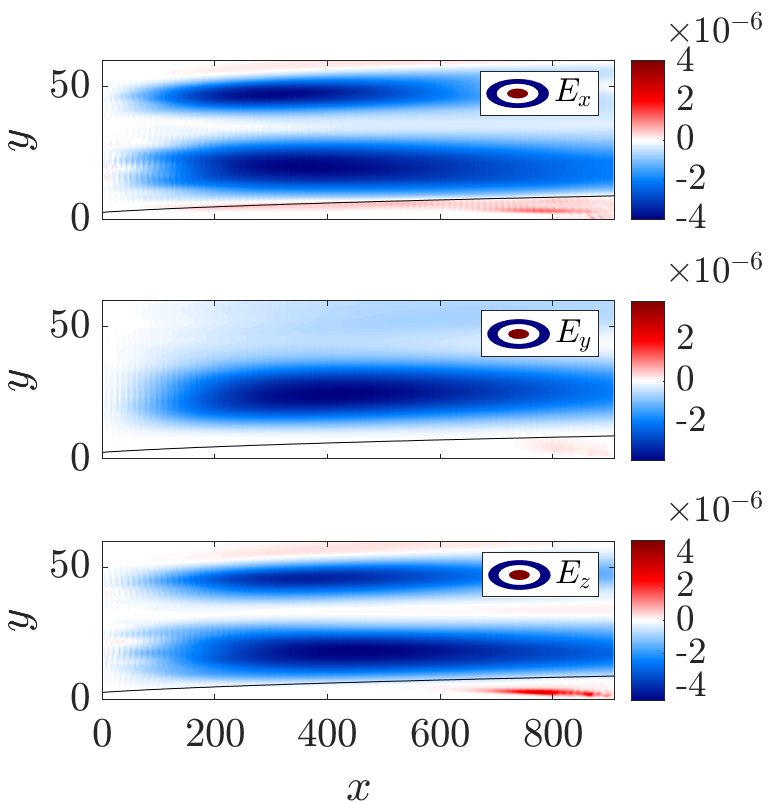}
		\caption{Non-linear component, $\mathbf{\Pi_N}$}
		\label{fig:Cnlin_Tu35_omega-0.1244_beta0.1257}
	\end{subfigure}
	\caption{Components of $\mathbf{P_Y}$, for $(\beta,\omega)=(0.126,-0.124)$ and $Tu=3.5\%$. Parameters: $\gamma = 0.929$; $\max\left(\left|\mathbf{\Pi_C}\right|\right) = \left[1.3 \times 10^{-6},2.6 \times 10^{-6},6.2 \times 10^{-7}\right]$.}
	\label{fig:Comp_Tu35_omega-0.1244_beta0.1257}
\end{figure}

The analysis of the PSD and its components for $(\beta,\omega)=(0.126,-0.124)$ and $Tu=0.5\%$, shown in figure \ref{fig:Comp_Tu05_omega-0.1244_beta0.1257}, brings interesting insights of the dynamics at higher frequencies. In this case, the structures are placed in the free stream, while little energy is present inside the boundary layer. Besides, the linear component, $\mathbf{\Pi_L}$, is more significant than the non-linear contribution, $\mathbf{\Pi_N}$, corroborating the behaviour described in section \ref{sec:spectra}.

These characteristics hold even when this same pair is considered for higher turbulent levels, as seen in figure \ref{fig:Comp_Tu35_omega-0.1244_beta0.1257}. At $Tu=3.5\%$, however, $\mathbf{\Pi_N}$ is proportionally stronger, rising above the magnitude of the correction component, $\mathbf{\Pi_C}$, and transfers energy out to other wavenumbers and/or frequencies. This explains the deviation from the purely linear growth, observed in figure \ref{fig:evo_Tu}, and displays the mechanism of the turbulent energy cascade acting on the free stream.

\section{Boundary layer structures} \label{sec:bl_struts}

\begin{figure}
	\centering
	\begin{subfigure}[b]{.49\textwidth}
		\centering
		\includegraphics[width=\linewidth]{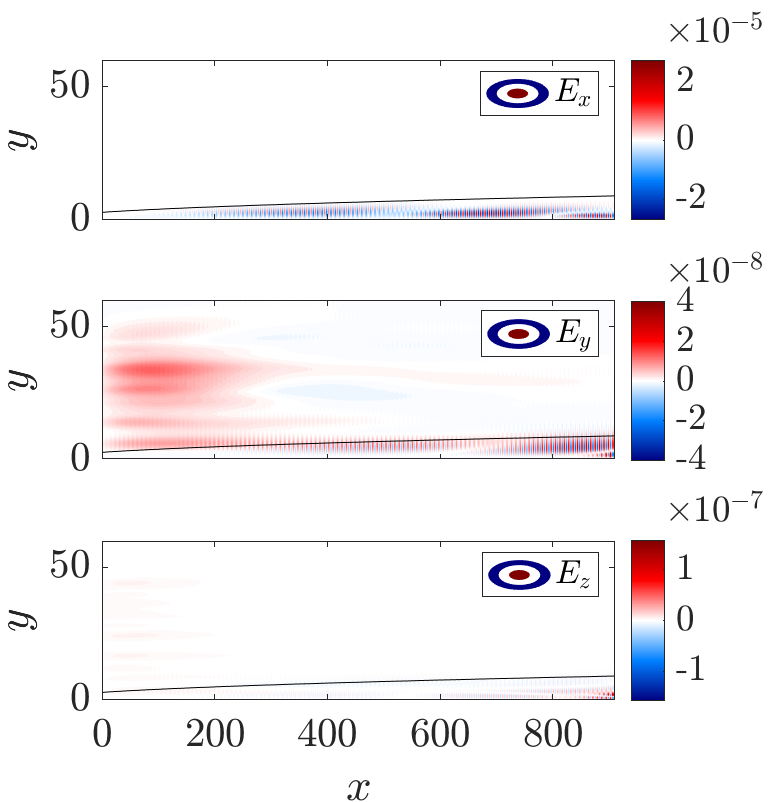}
		\caption{Linear component, $\mathbf{\Pi_{L}}$}
		\label{fig:Clin_Tu35_omega-0.0033_beta0.3770}
	\end{subfigure}%
	\begin{subfigure}[b]{.49\textwidth}
		\centering
		\includegraphics[width=\linewidth]{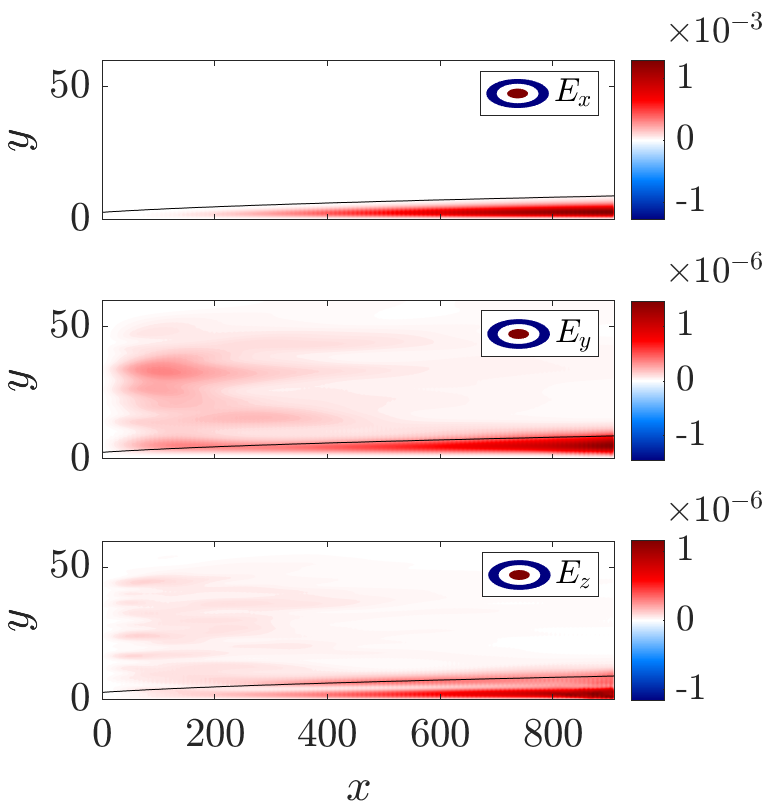}
		\caption{Non-linear component, $\mathbf{\Pi_N}$}
		\label{fig:Cnlin_Tu35_omega-0.0033_beta0.3770}
	\end{subfigure}
	\caption{Components of $\mathbf{P_Y}$, for $(\beta,\omega)=(0.377,-0.003)$ and $Tu=3.5\%$ (see Appendix \ref{sec:compPSD}). Note the difference in scales: the non-linear response is dominant while the linear one is negligible, having the same magnitude of the correction component. Parameters: $\gamma = 1.018$; $\max\left(\left|\mathbf{\Pi_C}\right|\right) = \left[6.4 \times 10^{-5},4.3 \times 10^{-8},1.3 \times 10^{-7}\right]$.}
	\label{fig:Comp_Tu35_omega-0.0033_beta0.3770}
\end{figure}

Linear and non-linear components of the PSD for $(\beta,\omega)=(0.377,-0.003)$ and $Tu=3.5\%$ are shown in figure \ref{fig:Comp_Tu35_omega-0.0033_beta0.3770}. Excluding the mean flow, this is the most energetic pair for all high FST levels, from $2.0\%$ and above. Contrary to the higher frequency pairs described in the previous section, here the energy is concentrated mainly in the boundary layer. The amplitude of structures is larger in the streamwise direction and the wave number $\beta$ matches the size of the streaky structures at the end of the physical domain, clearly observed in the snapshot of figure \ref{fig:snapshot}.

The prominence of the non-linear component, $\mathbf{\Pi_N}$, over the other two, $\mathbf{\Pi_{L}}$ and $\mathbf{\Pi_{C}}$, agrees with the behaviour shown in figure \ref{fig:evo_Tu}. It implies that most energetic structures of the flow, in higher FST levels, are mainly the product of the continuous non-linear forcing while displaying very little sensitivity to the linear interaction with the incoming turbulence via an initial condition at the intake.

\begin{figure}
	\centering
	\begin{subfigure}[b]{.49\textwidth}
		\centering
		\includegraphics[width=\linewidth]{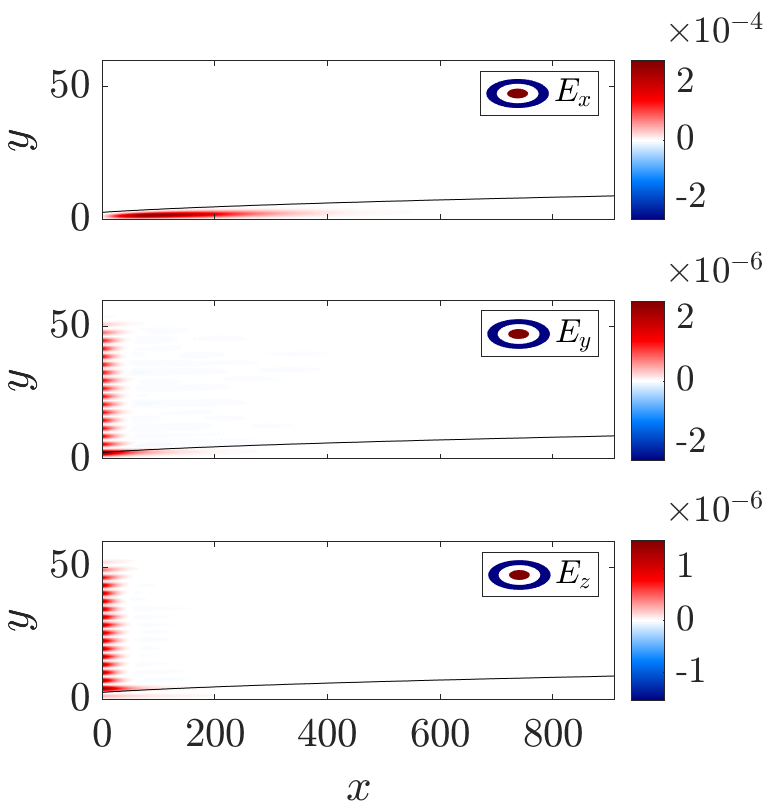}
		\caption{Linear component, $\mathbf{\Pi_{L}}$}
		\label{fig:LinCompTu35}
	\end{subfigure}%
	\begin{subfigure}[b]{.49\textwidth}
		\centering
		\includegraphics[width=\linewidth]{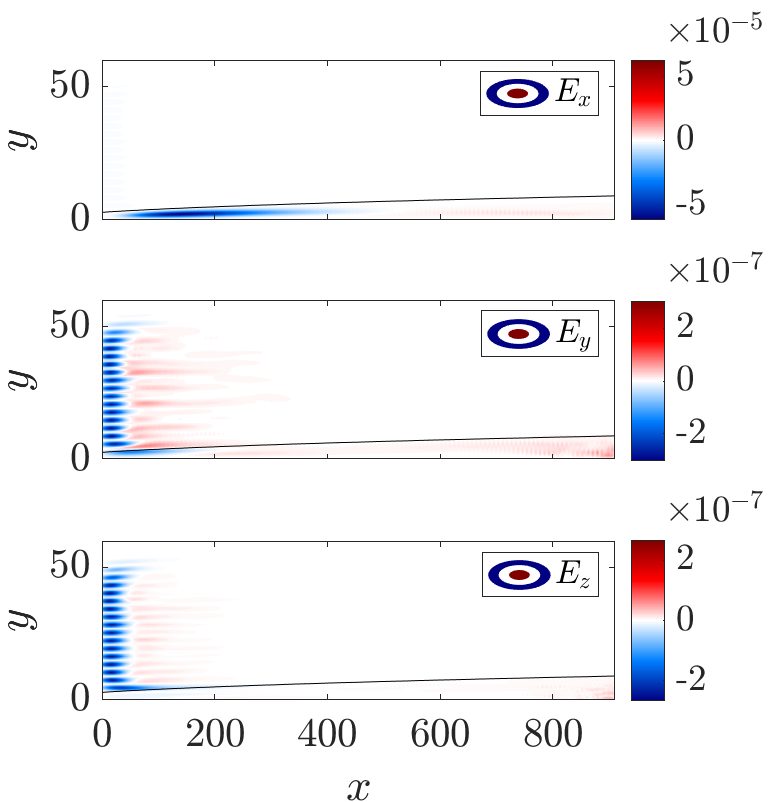}
		\caption{Non-linear component, $\mathbf{\Pi_N}$}
		\label{fig:NLinCompTu35}
	\end{subfigure}
	\caption{Components of $\mathbf{P_Y}$, for $(\beta,\omega)=(1.131,0.000)$ and $Tu=3.5\%$, (see Appendix \ref{sec:compPSD}). The linear response is the most important but the non-linear is non-negligible face to the correction component. Parameters: $\gamma = 0.989$; $\max\left(\left|\mathbf{\Pi_C}\right|\right) = \left[2.0 \times 10^{-6},1.3 \times 10^{-9},5.8 \times 10^{-10}\right]$.}
	\label{fig:Comp_Tu35_omega0.0008_beta1.1310}
\end{figure}

\begin{figure}
	\centering
	\begin{subfigure}[b]{.49\textwidth}
		\centering
		\includegraphics[width=\linewidth]{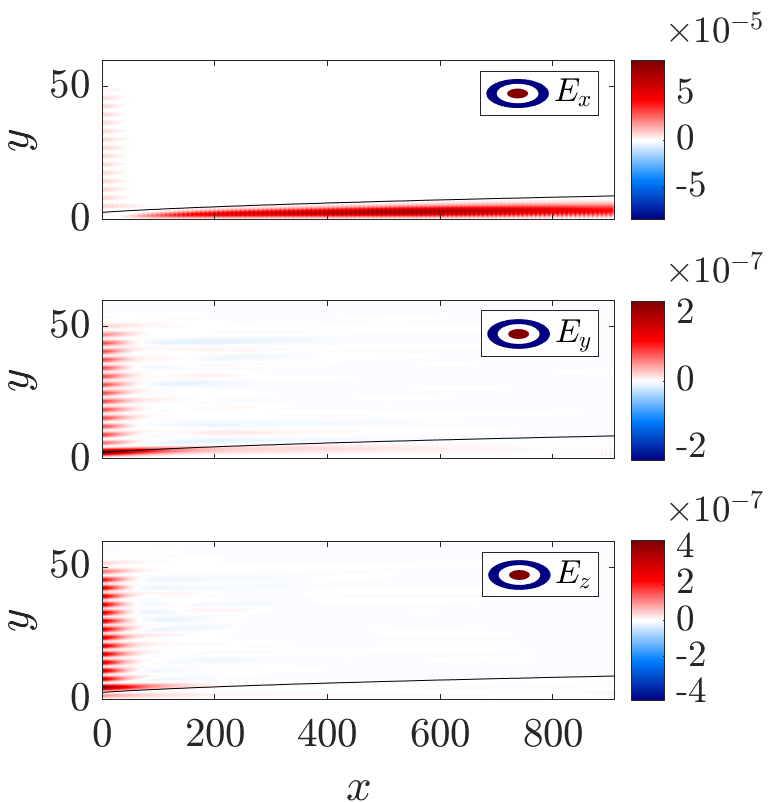}
		\caption{Linear component, $\mathbf{\Pi_{L}}$}
		\label{fig:Clin_Tu35_omega0.0000_beta0.5027}
	\end{subfigure}%
	\begin{subfigure}[b]{.49\textwidth}
		\centering
		\includegraphics[width=\linewidth]{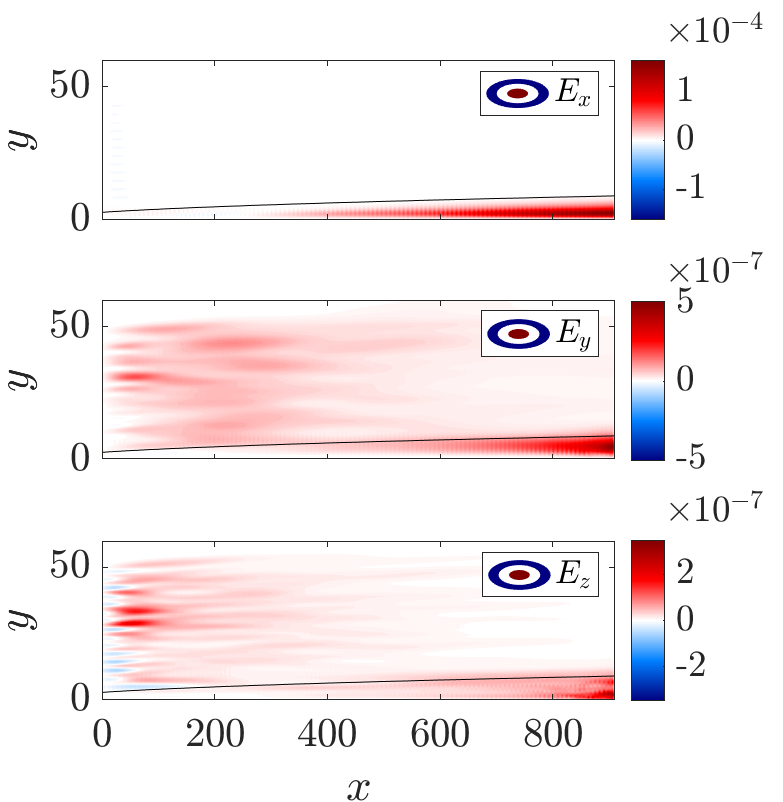}
		\caption{Non-linear component, $\mathbf{\Pi_N}$}
		\label{fig:Cnlin_Tu35_omega0.0000_beta0.5027}
	\end{subfigure}
	\caption{Components of $\mathbf{P_Y}$, for $(\beta,\omega)=(0.503,0.000)$ and $Tu=3.5\%$. Linear and non-linear responses have approximately the same order of magnitude. Parameters: $\gamma = 1.052$; $\max\left(\left|\mathbf{\Pi_C}\right|\right) = \left[8.0 \times 10^{-5},4.3 \times 10^{-8},2.2 \times 10^{-8}\right]$.}
	\label{fig:Comp_Tu35_omega0.0008_beta0.5027}
\end{figure}

Indeed, the streamwise elongated structures near the intake have a smaller spacing in $z$, suggesting a higher characteristic wavenumber $\beta$. In the $E_{L,in}$ spectrum of the $Tu=3.5\%$ case (figure \ref{fig:spectraComp}), this description is met by a peak at $(\beta,\omega)=(1.131,0.000)$. The reconstructed PSD components for this pair, shown in figure \ref{fig:Comp_Tu35_omega0.0008_beta1.1310}, indicate quite different dynamics from the previous analysis: linear excitation is predominant. For these structures, the non-linear response, smaller in magnitude, is still significant when compared to the correction, indicating a relevant outwards non-linear energy transfer flow, an effect that is especially strong for the $u$ component. Linear structures are most energetic in the streamwise direction and grow primarily in the upstream region of the domain, $x \in [0,300]$.

There exists still a third peak in the $E_{L,in}$ spectrum with intermediate wavenumber at $(\beta,\omega)=(0.503,0.000)$. In figure \ref{fig:Comp_Tu35_omega0.0008_beta0.5027} we observe that, as previously noted, dynamics at this pair are dominated by structures inside the boundary layer. Nevertheless, in contrast to the other two cases, linear and non-linear energy components have similar magnitudes. The linear component \revTwo{in the streamwise direction reaches its maximum} in the middle range or the domain $x \in [400,600]$, while the non-linear component is most important in more downstream positions. This constitutes a hybrid between the last two described cases, even though conclusions are not as robust since the correction component has comparable magnitude with the other two in the $u$ direction, violating the restrictions defined in section \ref{sec:rec_inputs}.

Thus, in summary, we evaluate three wavenumber-frequency pairs related to boundary layer streaks. For higher $\beta$ we observe mostly linear receptivity in the upstream part of the domain, as the linear component dominates the reconstructed PSD. Lower $\beta$ is characterised by a predominant non-linear receptivity, leading to downstream streaks of high energy. Intermediate wavenumbers display a transitional behaviour, with similar contributions of linear and non-linear receptivity mechanisms.

\section{Modal decomposition}

Up to this point, the available data was analysed from the energy point of view. Now, using modal decomposition techniques over the results of the spectral analysis performed in sections \ref{sec:freestream} and \ref{sec:bl_struts}, we are able to characterise, in terms of actual velocity fields, the most energetic structures and their related non-linear forcing, if relevant.

\subsection{Coherent structures generated by a linear mechanism}

\begin{figure}
	\centering
	\begin{subfigure}[b]{.47\textwidth}
		\centering
		\includegraphics[width=\linewidth]{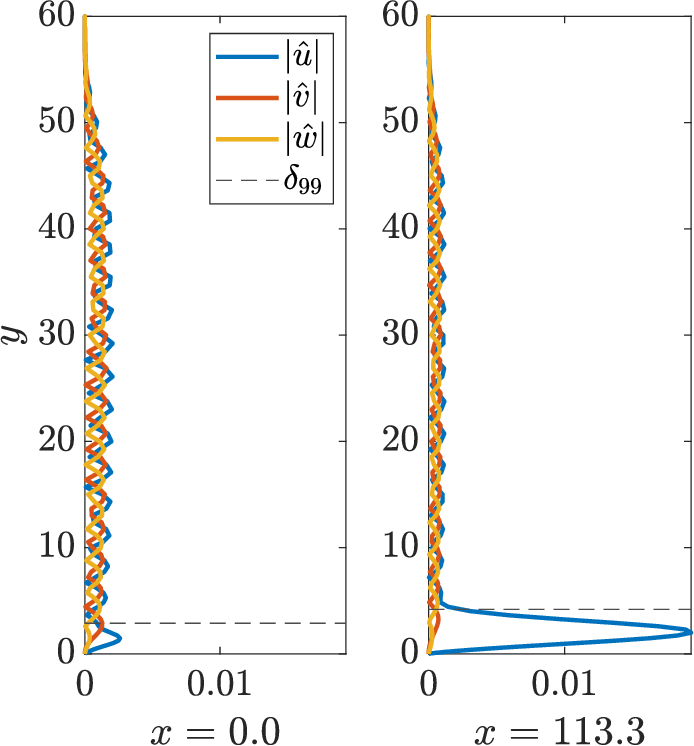}
		\caption{Amplitudes}
		\label{fig:SPOD_lin_streak_profile}
	\end{subfigure}%
	\begin{subfigure}[b]{.53\textwidth}
		\centering
		\includegraphics[width=\linewidth]{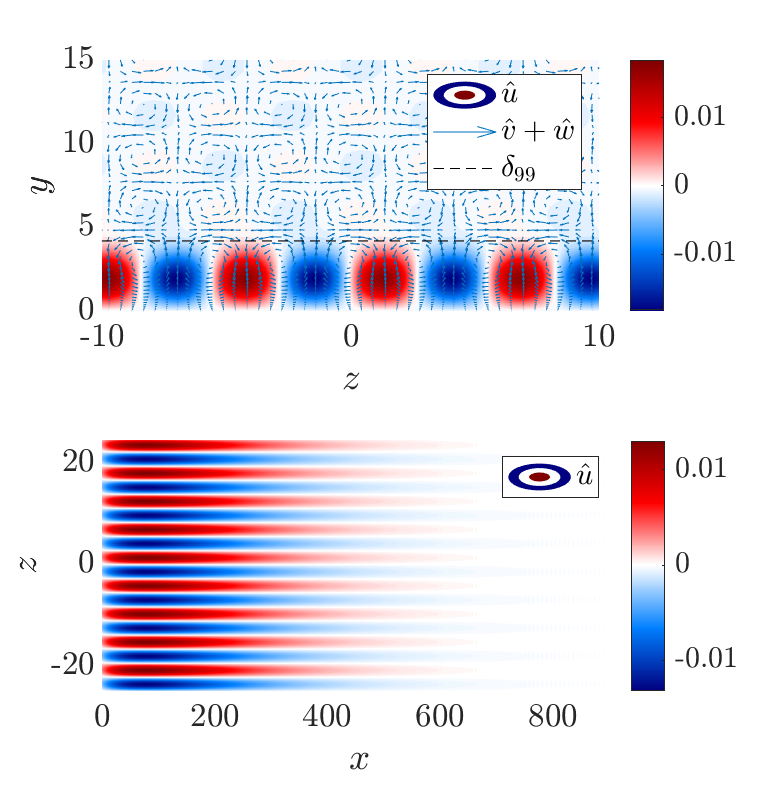}
		\caption{Spectral POD mode}
		\label{fig:SPOD_lin_streak}
	\end{subfigure}
	\caption{First spectral POD mode for $(\beta,\omega) = (1.131,0.000)$ at $Tu=3.5\%$, scaled by the respective eigenvalue. (a) Velocity profile of leading spectral POD mode at inlet and peak amplitude positions. (b) Real part of the leading spectral POD mode; (Top) Cross section at $x = 80$; (Bottom) Slice at $y = 1$, inside the boundary layer.}
	\label{fig:spectralPOD_lin_streak}
\end{figure}

\begin{figure}
	\centering
	\includegraphics[width=0.8\linewidth]{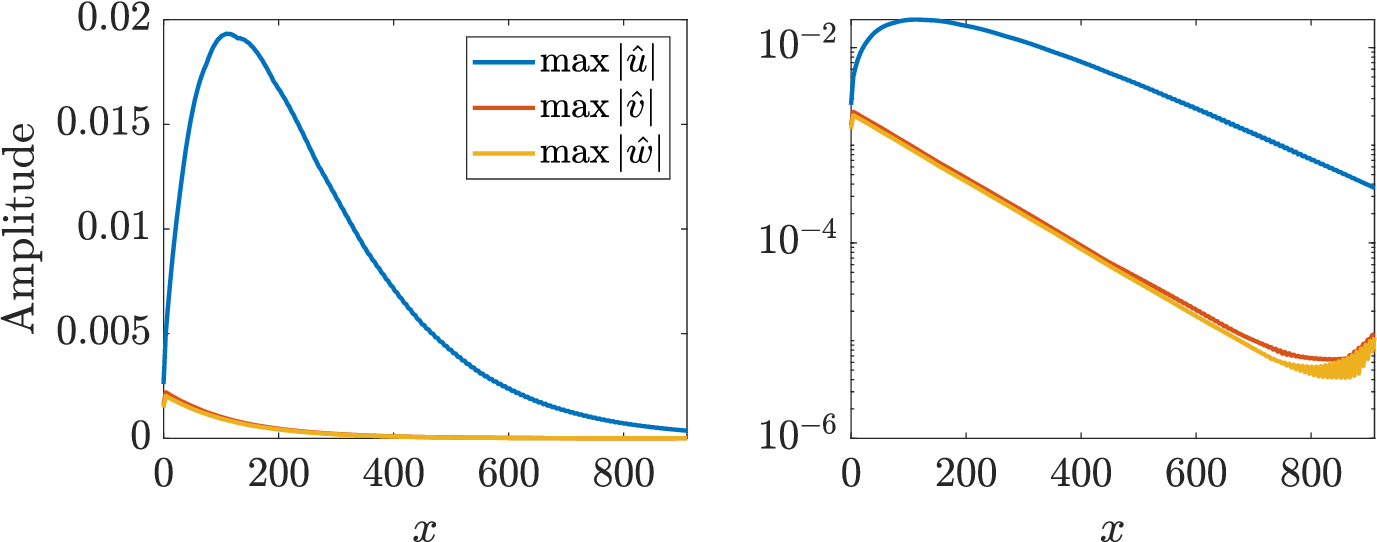}
	\caption{Maximum velocity amplitudes of the scaled spectral POD mode for $(\beta,\omega)=(1.131,0.000)$ at $Tu=3.5\%$. Transient growth of streaks generated by the linear mechanism (streamwise velocity component $u$), with streamwise vortices (spanwise and vertical components $v$ and $w$) that spatially decay.}
	\label{fig:transient_growth}
\end{figure}

First, we focus on the pair $(\beta,\omega) = (1.131,0.000)$ at $Tu=3.5\%$ and compute spectral POD modes of the data matrix $\mathbf{\hat{Y}}_L$, defined in eq. (\ref{eq:total_output}). The resulting leading mode, displayed in figure \ref{fig:spectralPOD_lin_streak}, features elongated streaky structures, with alternating regions of positive and negative streamwise velocity inside the boundary layer, intercalated by counter-rotating vortices bringing high-speed flow towards the boundary layer and ejecting low-speed flow from it, in a clear instance of the lift-up effect. 

Spatial transient growth is clear in figure \ref{fig:transient_growth}, which displays the maximum magnitudes of each velocity component at each streamwise position. While both spanwise and vertical components only decay, the streamwise component grows before exponentially decaying. Since these structures are spatially stable and only active in upstream positions, they cannot trigger the transition to turbulence in the present simulations, even though they might contribute to it through non-linear energy transfers.

\begin{figure}
	\centering
	\begin{subfigure}[b]{.5\textwidth}
		\centering
		\includegraphics[width=\linewidth]{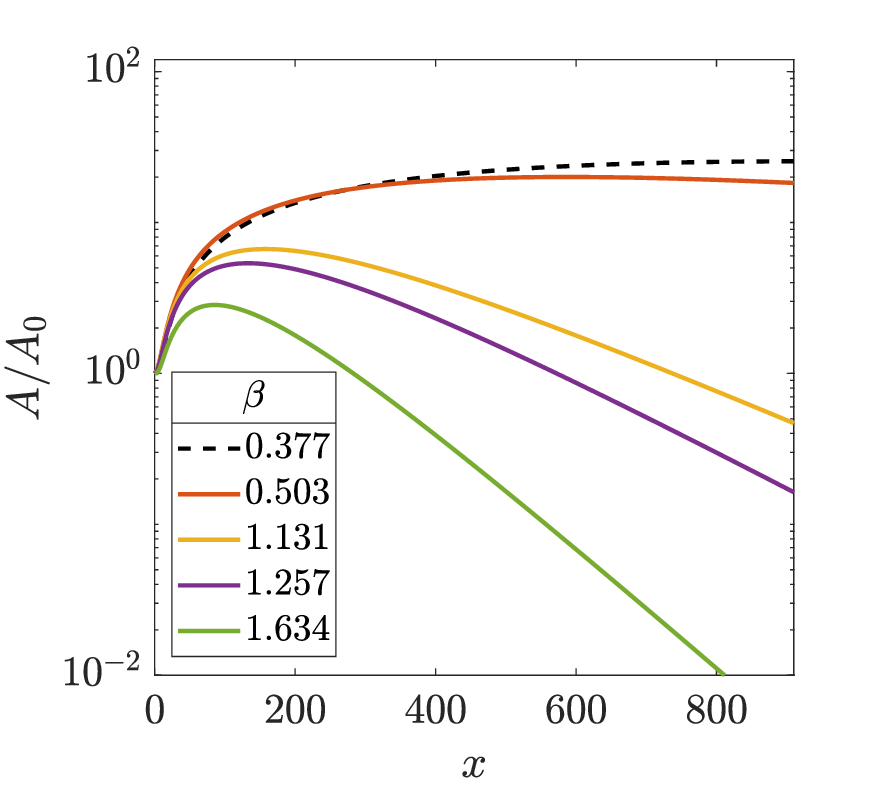}
		\caption{Optimal growth}
		\label{fig:optimal_linear}
	\end{subfigure}%
	\begin{subfigure}[b]{.5\textwidth}
		\centering
		\includegraphics[width=\linewidth]{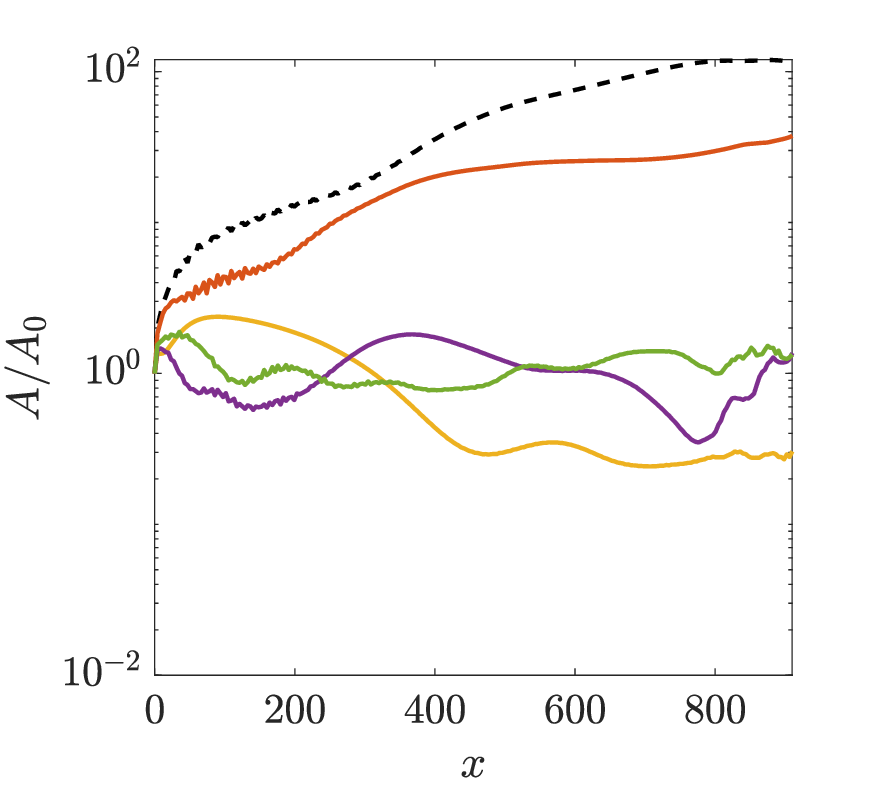}
		\caption{Measured amplifications}
		\label{fig:data_amps}
	\end{subfigure}
	\caption{\revTwo{Amplification of incoming perturbations. Amplitude is defined as the square root of the kinetic energy at a given position $x$. (a) Optimal growth according to \cite{doi:10.1063/1.869908} at spanwise wavenumbers corresponding incoming FST perturbations, computed for $\omega = 0$; (b) Amplifications measured for $Tu=3.5\%$. For $\beta=0.377$, not present in the OSS spectrum, the maximum optimal growth amplification is around 25, while the maximum measured amplification reaches 120.}}
	\label{fig:linear_amps}
\end{figure}

This spatial stability can be explained through transient growth theory. When the spanwise wavenumber, $\beta = 1.131$, at zero frequency, is introduced in the formulation presented in \cite{doi:10.1063/1.869908}, we conclude that optimal perturbations reach a maximum amplification before decaying in the streamwise direction. In a more complete analysis, we confirm that this is the case for all spanwise wavenumbers present in the FST spectrum at near-zero frequencies, as seen in figure \ref{fig:optimal_linear}. The largest linear amplification is found to be at the parameter corresponding to the structures previously shown in figure \ref{fig:Comp_Tu35_omega0.0008_beta0.5027}, which are, nevertheless, still less energetic than the ones presented in figure \ref{fig:Comp_Tu35_omega-0.0033_beta0.3770}.

It is worth noting that, since we can only introduce weak perturbations inside the boundary layer at the intake, which are not optimal in generating streaks through the linear amplification mechanism, the actual observed amplifications, shown in figure \ref{fig:data_amps}, are weaker than those predicted by the optimal growth theory for cases displaying linear receptivity, but larger for cases where non-linear interactions are important, such as $\beta = 0.503$ (figure \ref{fig:Comp_Tu35_omega0.0008_beta0.5027}) and $\beta=0.377$ (figure \ref{fig:Comp_Tu35_omega-0.0033_beta0.3770}, \revTwo{wavenumber not present in the OSS spectrum and weak in the FST energy spectrum at the intake, as seen in Appendix \ref{sec:propPert}}).

\subsection{Non-linear coherent structures}

Next, we focus on the pair $(\beta,\omega) = (0.377,-0.003)$ at $Tu=3.5\%$ and follow the procedure described in section \ref{sec:respectral POD}, to define an augmented state $\mathbf{\hat{Q}}$, composed only by the non-linear contributions of the model, which were identified to be the most important in this case. According to the notation of eq. (\ref{eq:total_output}), we have
\begin{equation} \label{eq:augmented_state}
	\mathbf{\hat{Q}} =
	\left[
	\begin{array}{c}
		\mathbf{\hat{Y}_N} \\
		\mathbf{\hat{F}}
	\end{array}
	\right]
\end{equation}
and, thus, spectral POD and forcing modes are linked by the relation $\Psi = \mathbf{R}\mathbf{B_f} \Phi$.

\begin{figure}
	\centering
	\begin{subfigure}[b]{.47\textwidth}
		\centering
		\includegraphics[width=\linewidth]{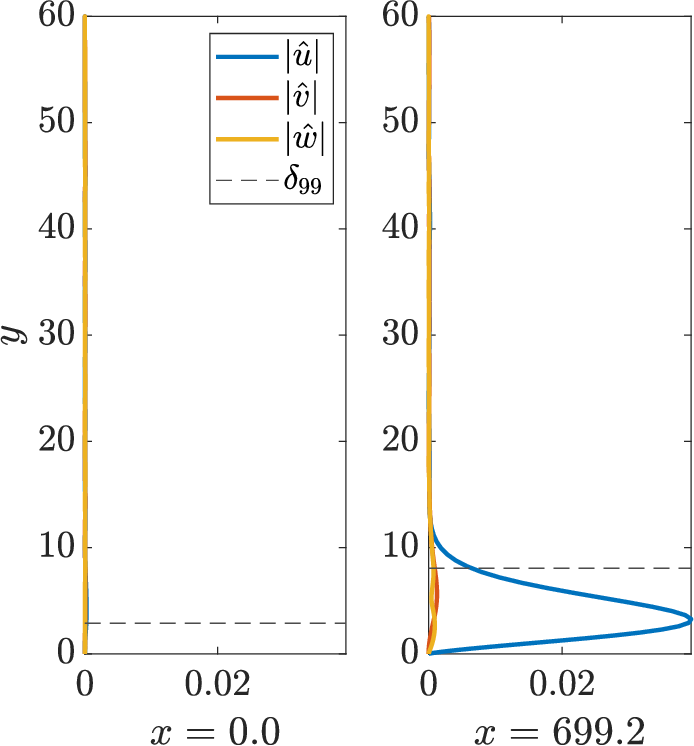}
		\caption{Amplitudes}
		\label{fig:SPOD_nonlin_streak_profile}
	\end{subfigure}%
	\begin{subfigure}[b]{.53\textwidth}
		\centering
		\includegraphics[width=\linewidth]{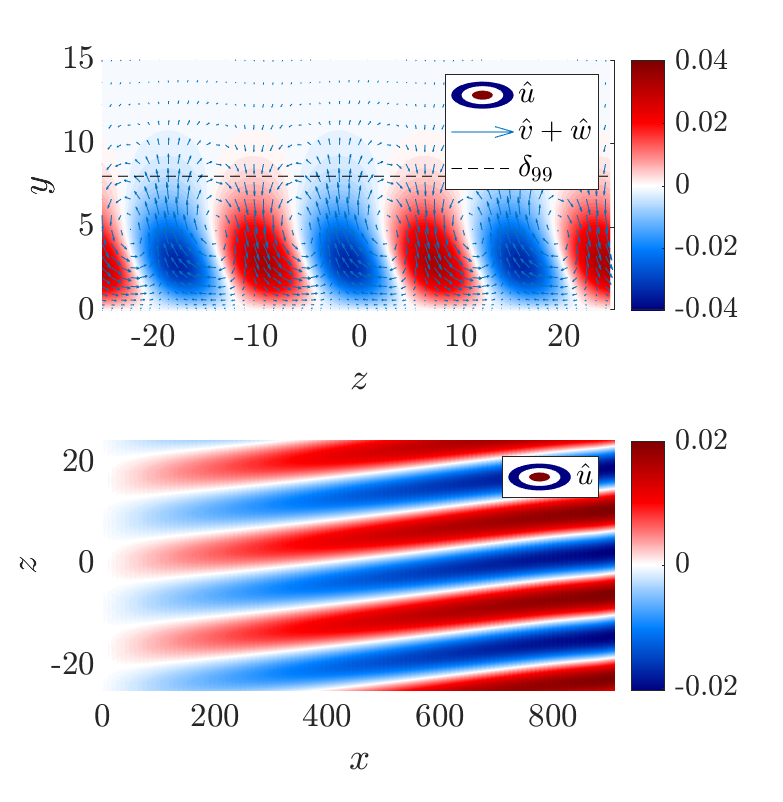}
		\caption{Spectral POD mode}
		\label{fig:SPOD_nonlin_streak}
	\end{subfigure}
	\caption{First spectral POD mode for $(\beta,\omega) = (0.377,-0.003)$ at $Tu=3.5\%$, scaled by the respective eigenvalue. (a) Velocity profile of leading spectral POD mode at inlet and position $x \approx 700$. (b) Real part of the leading spectral POD mode; (Top) Cross section at $x = 700$; (Bottom) Slice at $y = 1$, inside the boundary layer.}
	\label{fig:spectral POD_nonlin_streak}
\end{figure}

The leading spectral POD mode, shown in figure \ref{fig:spectral POD_nonlin_streak}, has the same overall shape found in streaks generated through linear receptivity: elongated structures, alternating streamwise velocity and counter-rotating vortices. However, significant velocity amplitudes are only present near the wall, below the position $y=15$. Besides, the characteristic wavenumber $\beta$ is smaller, such that the spacing between alternating regions is larger. As the frequency of this mode is not zero, the streaks appear inclined due to the perceived phase velocity; a similar mode is obtained for negative wavenumber, with mirrored inclination. These structures are the most energetic in cases where $Tu \geq 2\%$.

\begin{figure}
	\centering
	\includegraphics[width=0.8\linewidth]{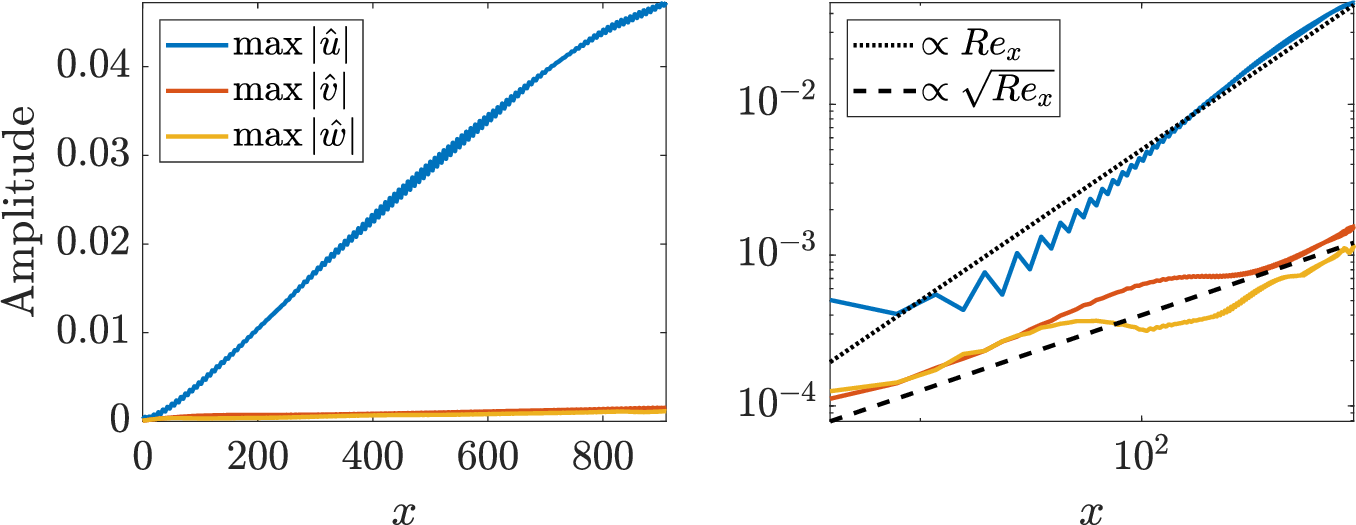}
	\caption{Maximum velocity amplitudes. Algebraic growth of streaks generated by the non-linear mechanism (streamwise velocity component $u$) with vortices (spanwise and vertical components, $v$ and $w$) growing proportional to $\sqrt{Re_x}$.}
	\label{fig:algebraic_growth}
\end{figure}

The streamwise evolution of streaks generated by the non-linear mechanism is not as steep as the one generated by linear growth. Streamwise velocity amplitudes are lower and quasi-streamwise vortices are weaker than those found at the intake for linear streaks. Contrary to their linear counterpart, the amplification is sustained over the whole length of the domain. Streamwise perturbations fit an algebraic growth pattern, while quasi-streamwise vortices scale proportionally to $\sqrt{Re_x}$ as seen in figure \ref{fig:algebraic_growth}. In practice, the non-linear interactions promote the necessary conditions to counteract the dampening effect of viscosity via a continuous forcing originating from the FST outside the boundary layer.

\begin{figure}
	\centering
	\includegraphics[width=0.8\linewidth]{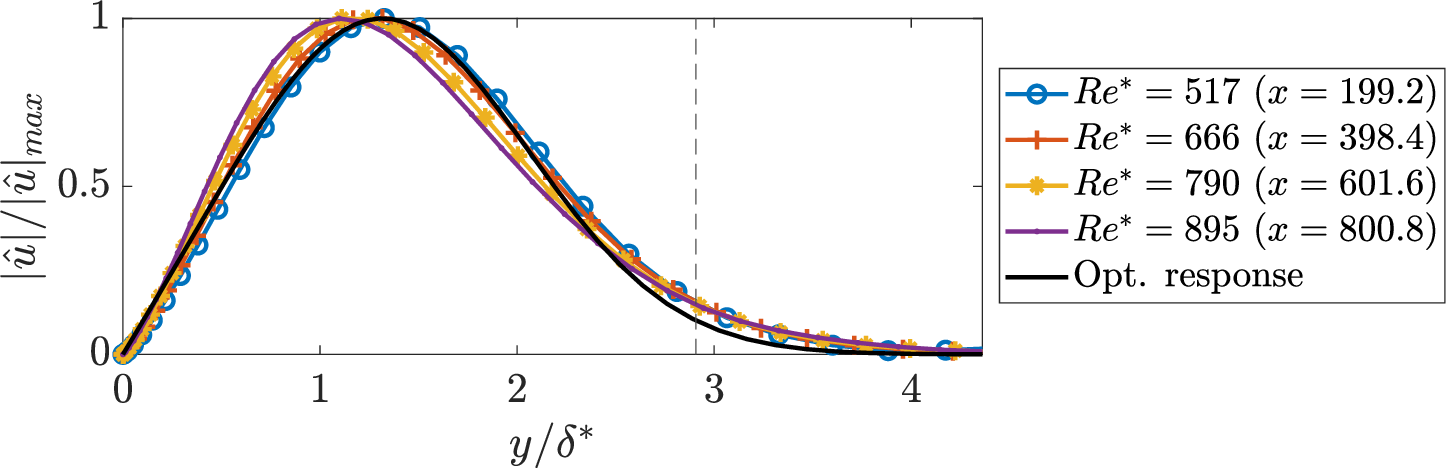}
	\caption{Comparison between optimal response profile, computed for $(\beta,\omega) = (0.377,0.000)$, and leading spectral POD mode. A vertical line is drawn at $\delta_{99}$.}
	\label{fig:opt_profile}
\end{figure}

As previously noted by \cite{sasaki_morra_cavalieri_hanifi_henningson_2020}, the velocity profile of streaks generated by the non-linear mechanism closely matches the optimal response of the linear amplification theory, computed for $(\beta,\omega) = (0.377,0.000)$ according to the procedure described in  \cite{doi:10.1063/1.869908}. Especially good agreement is achieved around $Re^*=600$, based on the boundary layer displacement thickness (see figure \ref{fig:opt_profile}). The fact that these streaks tend to conform to the same overall shape predicted by an optimal linear mechanism and, in turn, match the experimental profiles in \cite{westin_boiko_klingmann_kozlov_alfredsson_1994} and \cite{matsubara2001disturbance}, corroborates the conjecture concerning the existence of a strong dynamical attractor capable of ``bringing near to itself the velocity profile under most initial conditions", as mentioned in \cite{luchini2000}. In practice, it indicates the impossibility of asserting the linear or non-linear nature of streaky perturbations based solely on measurements of the corresponding velocity profiles at a given position.

\begin{figure}
	\centering
	\begin{subfigure}[b]{.47\textwidth}
		\centering
		\includegraphics[width=\linewidth]{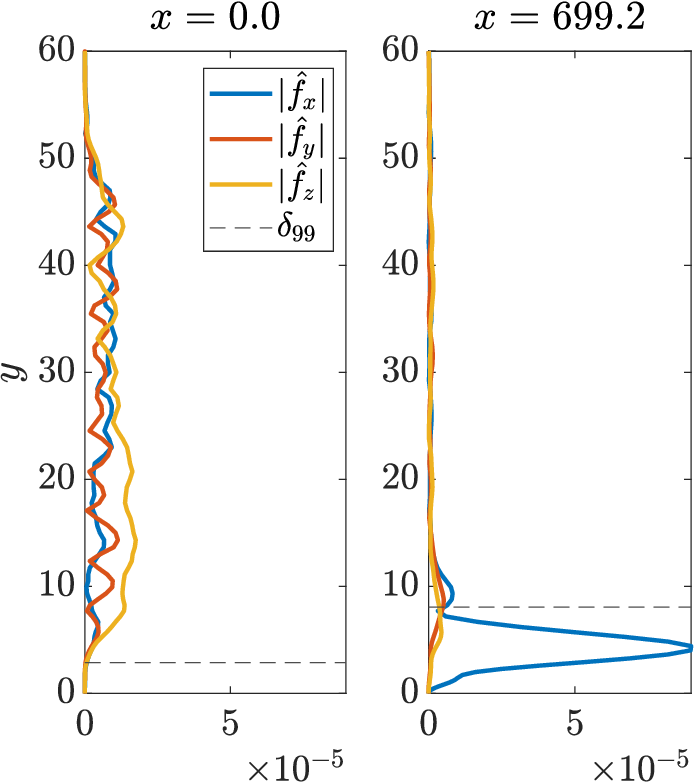}
		\caption{Forcing amplitudes}
		\label{fig:Forcing_nonlin_streak_profile}
	\end{subfigure}%
	\begin{subfigure}[b]{.53\textwidth}
		\centering
		\includegraphics[width=\linewidth]{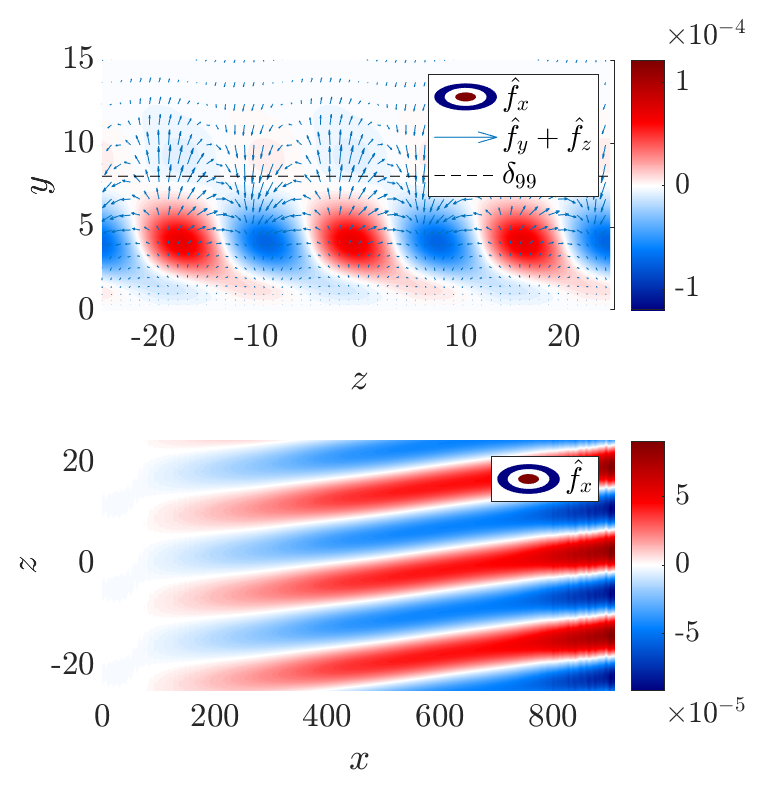}
		\caption{Forcing mode}
		\label{fig:Forcing_nonlin_streak_YZ_x700}
	\end{subfigure}
	\caption{Forcing mode for $(\beta,\omega) = (0.377,-0.003)$ at $Tu=3.5\%$, scaled by respective eigenvalue. (a) Forcing profile of leading mode at the inlet and at $x \approx 700$. (b) Real part of the leading forcing mode; (Top) Cross section at $x \approx 700$; (Bottom) Slice at $y = 3$, inside the boundary layer.}
	\label{fig:Forcing_nonlin_streak}
\end{figure}

In a last analysis, the shape of non-linear interactions can be analysed by looking at the first forcing mode, shown in figure \ref{fig:Forcing_nonlin_streak_YZ_x700}. From the RESPOD formulation, spectral POD and forcing modes are phase synchronized, such that by superposing the spectral POD mode shown in figure \ref{fig:spectral POD_nonlin_streak}, we observe that the non-linear forcing acts by feeding streamwise vortices just outside the edge of the boundary layer, while directly weakening the streaks inside of it.

This effect can be better described by first decomposing the non-linear forcing data into its components in each spatial direction and then reapplying the RESPOD analysis. Following the notation adopted in eq. (\ref{eq:augmented_state}), we construct
\begin{equation}
	\mathbf{\hat{F}_1} = \left[
	\begin{array}{l}
		\mathbf{\hat{F}_x}\\
		\emptyset\\
		\emptyset
	\end{array}\right], \quad
	\mathbf{\hat{F}_2} = \left[
	\begin{array}{l}
		\emptyset\\
		\mathbf{\hat{F}_y}\\
		\emptyset
	\end{array}\right], \quad
	\mathbf{\hat{F}_3} = \left[
	\begin{array}{l}
		\emptyset\\
		\emptyset\\
		\mathbf{\hat{F}_z}
	\end{array}\right], \quad
\end{equation}
such that
\begin{equation}
	\left\{
	\begin{array}{c}
		\mathbf{\hat{Y}_1} = \mathbf{R}\mathbf{B_f} \mathbf{\hat{F}_1} \\
		\mathbf{\hat{Y}_2} = \mathbf{R}\mathbf{B_f} \mathbf{\hat{F}_2} \\
		\mathbf{\hat{Y}_3} = \mathbf{R}\mathbf{B_f} \mathbf{\hat{F}_3}
	\end{array}
	\right., \quad
	\mathbf{\hat{Y}_N} = \mathbf{\hat{Y}_1} + \mathbf{\hat{Y}_2} + \mathbf{\hat{Y}_3},
\end{equation}
in order to obtain the component-wise augmented state $\mathbf{\hat{Q}_{a}}$, for which eq. (\ref{eq:snapshot_res}) gives the component-wise response and forcing modes,
\begin{equation} \label{eq:compwise_respod}
	\mathbf{\hat{Q}_{a}} =
	\left[
	\begin{array}{c}
		\mathbf{\hat{Y}_1} \\
		\mathbf{\hat{Y}_2} \\
		\mathbf{\hat{Y}_3} \\
		\mathbf{\hat{F}_1} \\
		\mathbf{\hat{F}_2} \\
		\mathbf{\hat{F}_3}
	\end{array}
	\right]
	\xrightarrow{\text{eq. (\ref{eq:snapshot_res})}}
	\mathbf{\tilde{\Psi}} =
	\left[
	\begin{array}{c}
		\mathbf{\Psi_1} \\
		\mathbf{\Psi_2} \\
		\mathbf{\Psi_3} \\
		\mathbf{\Phi_1} \\
		\mathbf{\Phi_2} \\
		\mathbf{\Phi_3}
	\end{array}
	\right]
	\implies
	\left\{
	\begin{array}{c}
		\mathbf{\Psi_1} = \mathbf{R}\mathbf{B_f} \mathbf{\Phi_1} \\
		\mathbf{\Psi_2} = \mathbf{R}\mathbf{B_f} \mathbf{\Phi_2} \\
		\mathbf{\Psi_3} = \mathbf{R}\mathbf{B_f} \mathbf{\Phi_3}
	\end{array}
	\right. .
\end{equation}

\begin{figure}
	\centering
	\begin{subfigure}[b]{.48\textwidth}
		\centering
		\includegraphics[width=0.95\linewidth]{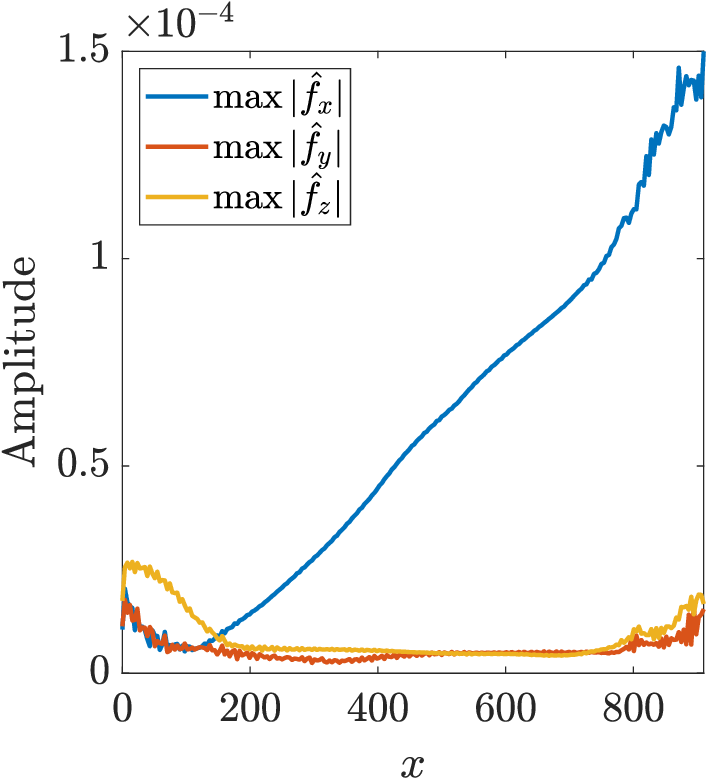}
		\caption{Maximum forcing amplitudes.}
		\label{fig:forcing_growth}
	\end{subfigure}%
	\begin{subfigure}[b]{.52\textwidth}
		\centering
		\includegraphics[width=0.95\linewidth]{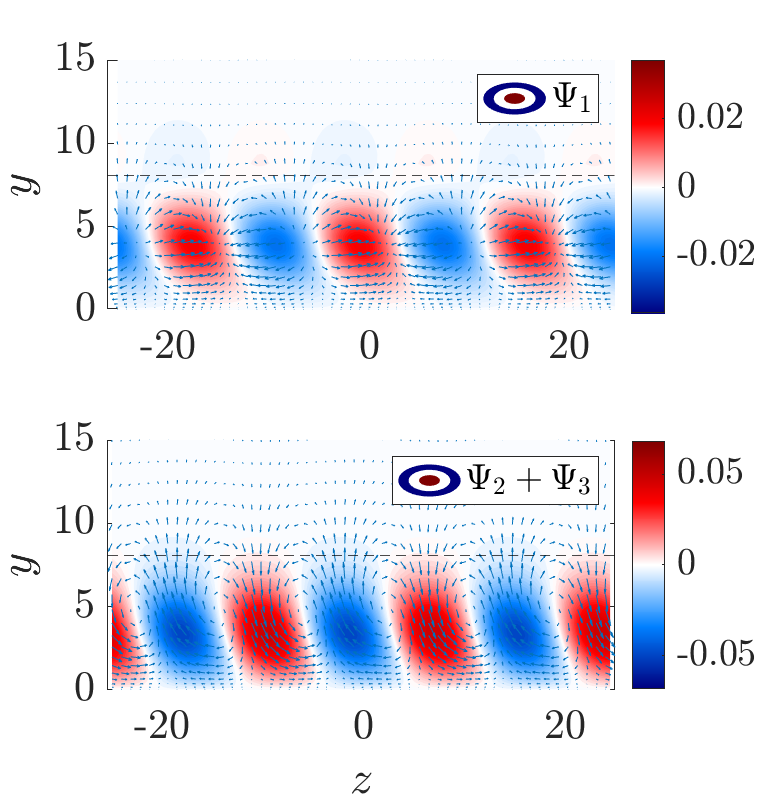}
		\caption{Response at $x=700$.}
		\label{fig:Compwise_RESPOD}
	\end{subfigure}
	\caption{Component-wise RESPOD analysis. (a) Maximum amplitudes of the forcing components along the streamwise direction. (b) Phase-synchronized responses due to the non-linear forcing. Horizontal line indicates $\delta_{99}$ thickness. (Top) Response generated by $\hat{f}_x$. (Bottom) Response generated by the composition of $\hat{f}_y+\hat{f}_z$.}
	\label{fig:compwise_analysis}
\end{figure}

In eq. (\ref{eq:compwise_respod}), the vectors $\mathbf{\Phi_1}$, $\mathbf{\Phi_2}$ and $\mathbf{\Phi_3}$, are respectively the separated components in $x$, $y$ and $z$ of the forcing mode presented in figure \ref{fig:Forcing_nonlin_streak}. Therefore, the vectors $\mathbf{\Psi_1}$, $\mathbf{\Psi_2}$ and $\mathbf{\Psi_3}$, displayed in figure \ref{fig:compwise_analysis}, are the corresponding phase-synchronized responses to each forcing component. Indeed, the results imply that the action of $\hat{f}_x$ generates streamwise structures acting in opposition of phase concerning the streaks that are mainly generated by $\hat{f}_y+\hat{f}_z$. This streamwise dampening effect is not enough to counteract streak growth, even though $\hat{f}_x$ steadily grows, reaching larger amplitudes than the other two components. This observation is supported by results from optimal growth theory, which indicate linear amplification from vortical, $\hat{f}_y$ and $\hat{f}_z$, forcing is stronger than from pure streamwise, $\hat{f}_x$, forcing.

Besides, the feature of opposing effects acting in a non-optimal manner to form the most energetic structures in the flow has already been observed in previous works \cite{nogueira_morra_martini_cavalieri_henningson_2021,morra_nogueira_cavalieri_henningson_2021}, where it was found that forcing fields computed from the non-linear terms of the Navier-Stokes equations tend to project poorly into the optimal input mode computed using resolvent analysis. In the present context, the results indicate that the streaks lose energy, through the $\hat{f}_x$ component, to other wavenumbers, in what could potentially be an initial stage of streak instability and breakdown \cite{hamilton_kim_waleffe_1995}.

\subsection{\revThree{Non-linear receptivity mechanism}} \label{sec:mechs}

Given the spanwise length of the domain, $L_z$, and $\beta_0 = 2 \pi / L_z$, streaks generated by the non-linear mechanism appear at approximately $(\beta,\omega) = (3 \beta_0,0)$. This seems to suggest, at least for the flow case considered here, a different receptivity mechanism than the classical oblique wave setup described in section \ref{sec:intro}, since the wavenumber in question cannot be reached by triadic interaction of the type $(\pm \beta,\omega) \rightarrow (2\beta,0)$ \cite{berlin_wiegel_henningson_1999,doi:10.1063/1.1456062}. This could, however, imply an interaction between oblique waves of different wavenumbers, such as $[(\beta,\omega),(-2\beta,\omega)] \rightarrow (3\beta,0)$.

A meaningful analysis of this mechanism would require a decomposition of the non-linear convection term into its triadic components in both spanwise wavenumber and frequency, as described in eq \ref{eq:conv_f}. \revThree{The identification of a set of triads linking a non-linear pair $(\beta,\omega)$ localised inside the boundary-layer to two linear pairs predominantly present in the free-stream would constitute a useful data-driven approach to identify non-linear receptivity mechanisms induced by FST in a statistically stationary setup. Moreover, the ranked non-linear forcing modes could be employed to characterise perturbation-perturbation interactions neglected in restricted non-linear models \cite{farrell_ioannou_jiménez_constantinou_lozano-durán_nikolaidis_2016}.}

This is not accomplished in the present work for two main reasons: (i) The databases were setup to resolve mainly low-frequency dynamics, only a small part of the full frequency spectrum of the incoming FST perturbations, as shown in Appendix \ref{sec:propPert}; (ii) The spectral decomposition of less energetic pairs $(\beta,\omega)$ inevitably encounters significant windowing correction components $\mathbf{\Pi_{C}}$, violating the rule established in section \ref{sec:rec_inputs}. Arguably, a triadic analysis could be performed with a properly time-resolved database.

\section{Conclusions} \label{sec:conclusions}

In the present study, we combined spectral estimation with the POD method and the resolvent analysis framework to distinguish linear and non-linear coherent structures present in simulations of transitional boundary layers over flat plates without leading edge, subject to multiple levels of free-stream turbulence (FST). This was accomplished with the employment of an input-output (state-space) formulation that segregates external turbulent forcing, acting in the fringe zone, from volumetric inputs computed directly from simulated fluctuation fields using the non-linear convection term, $f_i = - u^\prime_j \frac{\partial u^\prime_i}{\partial x_j}$.

At first, the analysis of the simulation's statistical power spectra showed that structures are amplified by the increased FST levels, $Tu$, mainly in the lower frequency range, defined at $|\omega| < 0.026$, a value found to be related to the incoming FST spectrum. In sequence, two main trends were identified by tracking the behaviour of the most energetic pairs $(\beta,\omega)$ at each $Tu$ level: while higher frequencies evolve with a scale closer to $Tu^2$, indicating a linear interaction with the incoming turbulent energy, lower frequencies display a steeper amplification, characteristic of non-linear mechanisms. These trends were once more verified by superposing all available power spectra in frequency. Again, higher frequencies collapse when normalised by $Tu^2$. Concurrently, lower frequencies scale with a factor closer to $Tu^4$ for the present numerical database. These scalings are consistent with linear and non-linear receptivity mechanisms, respectively.

Once we computed the reconstructed spectral response of the system through the input-output formulation, we integrated the energies of linear and non-linear response components in two distinct regions, inside and outside the Blasius boundary layer. With this, lower frequency energy peaks were linked to boundary layer structures, while higher frequency peaks were established to be the result of the incoming turbulent flow.

In the free stream, the peaks in the linear component spectrum often translate to negative non-linear contributions, a feature attributed to the mixing and redistributing properties, between triads of wavenumbers and frequencies, of the turbulent energy cascade. On the other hand, the kinetic energy inside the boundary layer is found primarily in the non-linear component spectrum, at $(\beta,\omega) = (0.377,-0.003)$, with less energetic peaks present in the lower frequency range of the linear component spectrum, especially at $(1.131,0.000)$.

The application of the spectral POD method over the data at $(\beta,\omega) = (1.131,0.000)$ for $Tu=3.5\%$ reveals dynamics dominated by streaky structures upstream, near the intake of the numerical domain. These are largely a result of the linear response of the system and display spatially stable spanwise and vertical velocity components, with strong amplification of the streamwise component, readily followed by an exponential decay, characteristics of transient growth. Thus, streaks generated by the linear mechanism do not contribute directly to transition in the present case.

When the RESPOD method is applied to the data at $(\beta,\omega) = (0.377,-0.003)$, however, quite different dynamics are unveiled: streaks are solely the result of the continuous non-linear forcing and, contrary to the transient dynamics observed before, are steadily amplified throughout the whole domain, along with vortices that grow proportionally to $\sqrt{Re_x}$.  The velocity profile of the leading mode, computed using only the non-linear component of the system's response matches the optimal amplification profile from transient growth theory \cite{luchini2000}, supporting the conjecture of a strong dynamical attractor within the boundary layer. The computed leading forcing mode for streaks generated by the non-linear mechanism reveals non-optimal amplification mechanisms, in the sense that the forcing acts both dampening and feeding streaks, a feature which could potentially indicate the beginning of streak breakdown \cite{hamilton_kim_waleffe_1995}. Also, the presence of streaks generated by the non-linear mechanism at $(\beta,\omega) = (0.377,-0.003) \approx (3 \beta_0,0)$ suggests a different mechanism from the classical oblique wave setup \cite{berlin_wiegel_henningson_1999,doi:10.1063/1.1456062}, at least in the considered flow case.

Arguably, the simulation setup studied is idealised and strong assumptions are made when constructing an incoming turbulent field with OSS modes on the continuous spectrum. In the presence of a leading edge, turbulence could be introduced inside the boundary layer near the stagnation point, greatly favouring the linear mechanism, which would result in an overall energy dependency of $E \propto Tu^2$, as measured by \cite{fransson_matsubara_alfredsson_2005}. Moreover, the identified non-linear mechanism could be important even in the case of a turbulent boundary layer, contributing to the regeneration cycle of turbulent streaks described in \cite{hamilton_kim_waleffe_1995} and \cite{brandt2014}. These considerations are, however, left open to future works.

The numerical methods devised in this manuscript allowed the identification of both linear and non-linear receptivity mechanisms in the early stages of transition and the description of the non-linear forcing capable of generating the identified most energetic structures in the flow. Other than simulation data, in the form of flow snapshots, the methodology requires the knowledge of the linear operators involved, as well as boundary conditions. In the presented workflow, the geometry and size of the numerical mesh made possible the construction of the linear operators and computation of the non-linear forcing term outside a numerical solver. This might not be the case for larger simulations and more complex geometries, for which the computation of the convective non-linear term and resulting linear and non-linear components of the full system response must be done employing the same operators implemented by the specific solver used to perform the simulations. \revIntern{In particular, one natural future development of the present work is the inclusion of a leading edge, which requires curvilinear meshes with corresponding spatial derivative operators, and different FST generation schemes to introduce perturbations upstream of the stagnation point, far from no-slip surfaces.} Therefore, we stress that the approach is general and could potentially be extended to any simulation for which receptivity to incoming perturbations needs to be assessed, contributing, in that sense, not only to the advancement of the research concerning the transition to turbulence but also to the field of non-linear dynamics as a whole.

\section*{Acknowledgements}
This work has been financially supported by Fundação de Amparo à Pesquisa do Estado de São Paulo, FAPESP, under grants nos. 2019/27655-3, 2020/14200-5 and 2022/01424-8. We would like to thank Eduardo Martini, Kenzo Sasaki and José Alarcón for the fruitful discussions related to the work developed here.

\section*{Declaration of interests}
The authors report no conflict of interest.

\appendix

\section{\revTwo{Properties of inflow perturbations}} \label{sec:propPert}

\revTwo{As described in section \ref{sec:fringe_forcing}, we introduce synthetic homogeneous FST into the simulation domain by forcing a set of OSS modes in the continuous spectrum branch inside the fringe region. In this section, the spectrum of OSS modes is presented and the homogeneity property of the FST is discussed.}

\begin{figure}
	\centering
	\begin{subfigure}[b]{.5\textwidth}
		\centering
		\includegraphics[width=\linewidth]{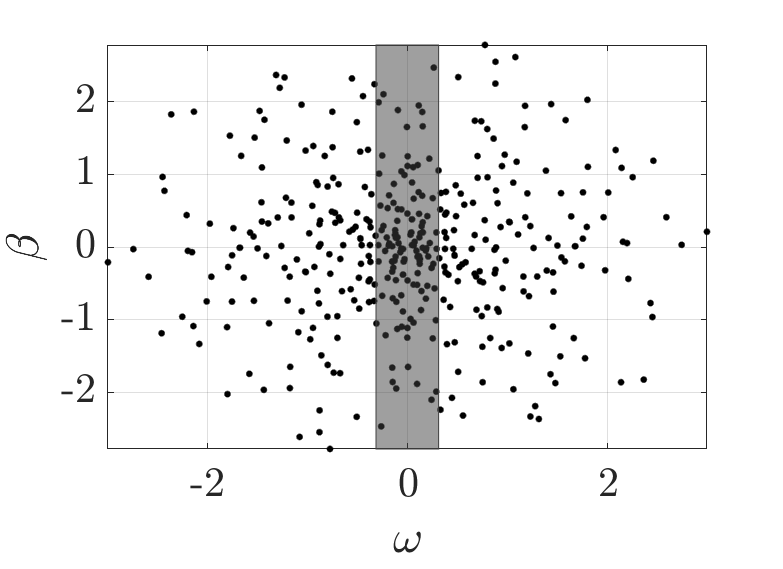}
		\caption{}
		\label{fig:OSS_spectrum}
	\end{subfigure}%
	\begin{subfigure}[b]{.5\textwidth}
		\centering
		\includegraphics[width=\linewidth]{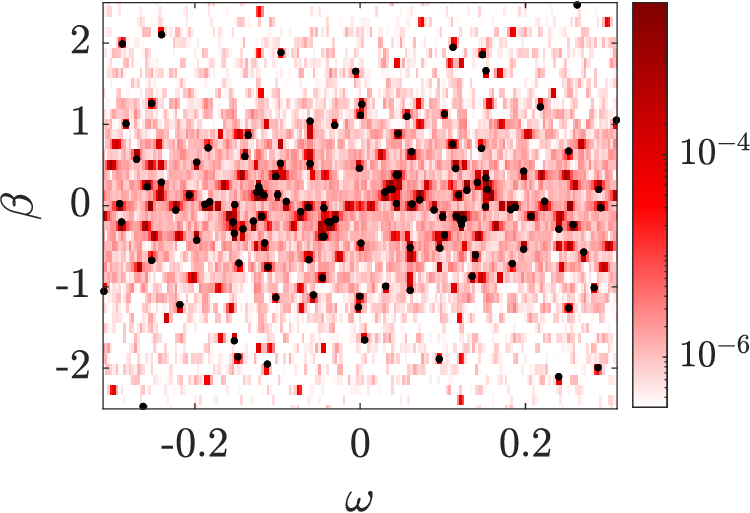}
		\caption{}
		\label{fig:specSup}
	\end{subfigure}
	\caption{Spectrum of perturbations introduced in the fringe zone. Fig. \ref{fig:OSS_spectrum}: Full spectrum of OSS modes where the grey band represents the frequencies resolved by the snapshots of simulations; Fig \ref{fig:specSup}: Measured spectrum at the inlet for the $Tu=3.5\%$ case, superposed by the OSS modes spectrum, with colours representing turbulent energy.}
	\label{fig:oss_spec}
\end{figure}

\revTwo{Figure \ref{fig:oss_spec} shows the spectrum of perturbations introduced in the fringe zone, as a function of spanwise wavenumber, $\beta$ and frequency, $\omega$. Since it is known that modes in the continuous branch have phase speed approximately equal to the free-stream velocity, $U_\infty$, we apply Taylor's hypothesis, $\omega = \alpha U_\infty$, to compute $\omega$ as a function of the computed streamwise wavenumber, $\alpha$, given by spatial stability. It should be noted in figure \ref{fig:OSS_spectrum} that the snapshots taken from the simulations, spaced by time steps of $\Delta t=10$ to capture the low-frequency dynamics of streaks in bypass transition, do not resolve the full perturbation spectrum in frequency.}

\revTwo{Methods to synthetically generate FST via OSS modes have found some criticism in the fluid mechanics community. Particularly in the work of \cite{dong_wu_2013}, it is argued that continuous OSS spectra might be unsuitable to characterise free-stream disturbances and their interaction with the boundary layer because of two main factors: the phenomenon labelled entanglement of Fourier modes and the observation that low-frequency disturbances appear to force preferentially the streamwise component of the fluctuations in the free-stream, in detriment of the transverse ones. Here, these concerns are addressed based on the statistical data from the inlet perturbations of the simulations considered in this work.}

\revOne{First, the entanglement of Fourier modes is a non-physical property arising from the parallel flow approximation of OSS equations, which potentially generates spurious perturbations if such modes are introduced as inlet conditions. There is, however, a distinction between this description and the approach employed in the present work, based on \cite{brandt_schlatter_henningson_2004}. Considering the momentum equations written in eq. (\ref{eq:ns_1}),}
\begin{equation}
	\left.
	\begin{aligned}
		\frac{\partial \mathbf{u^\prime}}{\partial t} = LNS (\mathbf{u^\prime},\mathbf{U_{BL}}) + \revOne{f(\mathbf{u^\prime})} + \sigma(x)(\mathbf{\zeta}-\mathbf{u^\prime}),& \\
		\nabla \cdot \mathbf{u^\prime} = 0,&
	\end{aligned}
	\right\}
\end{equation}
\revOne{the fringe $\sigma(x)(\mathbf{\zeta}-\mathbf{u^\prime})$ term acts as a proportional controller that imposes a body force capable of bringing the flow near to the desired state introduced by the forcing term $\mathbf{\zeta}$ composed of a superposition of OSS modes. Therefore, $\mathbf{\zeta}$ is not imposed directly and the state inside the fringe is always a solution of an externally forced incompressible NS system, for which no parallel flow assumptions are made, rather than the solution of the OSS equations. This effect can be observed in figure \ref{fig:specSup}, where the energy spectrum at the inlet is superposed by the OSS modes spectrum. Through careful design of the fringe region, we can match energy peaks with the location of OSS modes, even though peaks are also present in different locations due to the influence of the NS system. For a more detailed description of the effects of the fringe parameters, the reader is referred to \cite{chevalier2007simson}.}

\begin{figure}
	\centering
	\begin{subfigure}[b]{.42\textwidth}
		\centering
		\includegraphics[width=\linewidth]{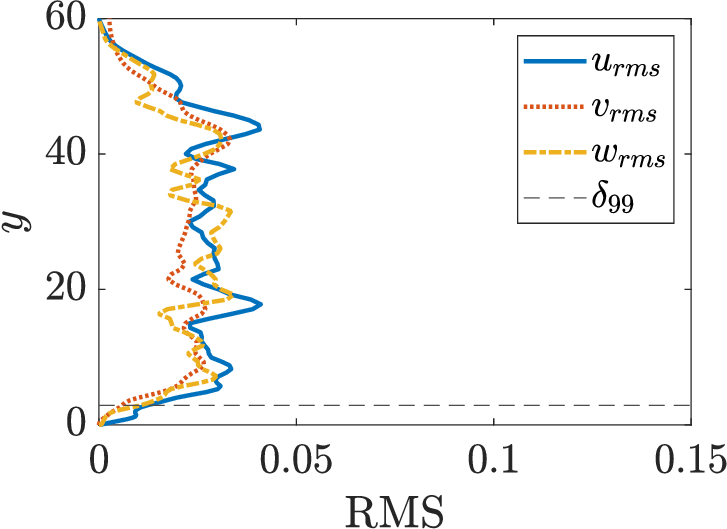}
		\caption{$x=0$}
		\label{fig:rms_inlet}
	\end{subfigure}%
	\begin{subfigure}[b]{.42\textwidth}
		\centering
		\includegraphics[width=\linewidth]{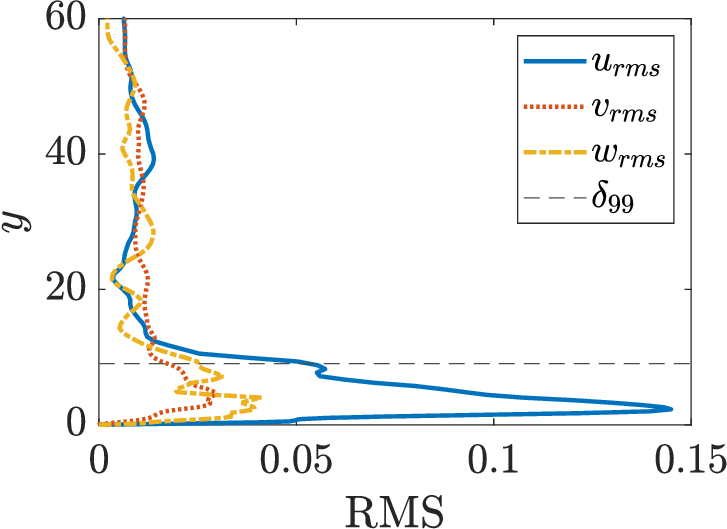}
		\caption{$x=910$}
		\label{fig:rms_end}
	\end{subfigure}
	\caption{Root mean squared value of the velocity fluctuations, averaged over span and time directions, for the case of $Tu=3.5\%$. Fig \ref{fig:rms_inlet}: RMS values at the intake; Fig \ref{fig:rms_end}: RMS values before the fringe.}
	\label{fig:rms_fluc}
\end{figure}

\revTwo{Second, the preferential amplification of streamwise the fluctuations in the free-stream observed by \cite{dong_wu_2013} in the context of OSS equations is not present in the simulations considered in this work. As shown in figure \ref{fig:rms_inlet}, the root-mean-squared (RMS) values for all three perturbation components have roughly the same magnitude at the inlet and are mainly located outside the boundary layer. The fringe region is capable of homogenising the streamwise-dominated perturbations present upstream of it, generated by the streaky dynamics of bypass transition, as seen in figure \ref{fig:rms_end}.}

\revTwo{Finally, one should note that, since we deal with input-output analysis in this work, the methods presented are agnostic to the type of perturbations introduced. In other words, even though some results might be influenced by the way synthetic turbulence is generated, the formulation is general enough and does not limit the application of different techniques for the generation of incoming perturbations, given that adequate adaptations are applied to the input-output system.}

\section{Linear operators, sparsity and boundary conditions} \label{sec:linOp}

The state-space formulation presented in eq. (\ref{eq:state_space}) is directly derived from eq. (\ref{eq:lin_base}). For a model discretised in $N$ spatial points and a base-flow vector
\begin{equation}
	\mathbf{U}= \left[
	\begin{array}{l}
		\mathcal{U}\\
		\mathcal{V}\\
		\emptyset
	\end{array}\right], \quad \mathcal{U},\mathcal{V},\emptyset \in \mathbb{R}^{N \times 1}
\end{equation}
composed of row-wise stacked components, the linearised Navier-Stokes operator, $\mathbf{L}$, is defined as
\begin{equation}\label{eq:L}
	\mathbf{L} = \left[
	\begin{array}{cccc}
		\mathbf{K} + \left(\mathbf{D}_x \mathcal{U}\right)^T\mathbf{I} & \left(\mathbf{D}_y \mathcal{U}\right)^T\mathbf{I} & \mathbf{Z} & \mathbf{D}_x \\
		\left(\mathbf{D}_x \mathcal{V}\right)^T\mathbf{I} & \mathbf{K} + \left(\mathbf{D}_y \mathcal{V}\right)^T\mathbf{I}	& \mathbf{Z} & \mathbf{D}_y \\
		\mathbf{Z} & \mathbf{Z} & \mathbf{K} & i \beta \mathbf{I} \\
		\mathbf{D}_x & \mathbf{D}_y & i \beta \mathbf{I} & \mathbf{Z}
	\end{array}
	\right]
\end{equation}
where
\begin{equation}\label{eq:K}
	\mathbf{K} =  \mathcal{U}^T \mathbf{D}_x + \mathcal{V}^T \mathbf{D}_y + \frac{1}{Re} \left(\mathbf{D}_{xx} + \mathbf{D}_{yy} - \beta^2 \mathbf{I} \right) + \mathbf{\sigma}^T \mathbf{I}
\end{equation}
and $\sigma \in \mathbb{R}^{N \times 1}$ is the fringe gain from figure \ref{fig:gain_fringe}. Matrices $\mathbf{I}$ and $\mathbf{Z}$ are identity and zero respectively. Matrices $\mathbf{D}_x$, $\mathbf{D}_y$ are first and $\mathbf{D}_{xx}$, $\mathbf{D}_{yy}$ are second spatial derivatives in the respective directions. The superscript $\{\cdot\}^T$ indicates transpose. All specified matrices have dimension $N \times N$.

Depending on the size of the model, matrices can be costly to store and manipulate. In this work, for instance, the 2D grid has a total of $N = 256 \times 121 = 30,976$ points. If values are stored in 16 bytes (real and imaginary parts as 8 bytes double precision floats each), $\mathbf{L}$ should amount to approximately 240 Gigabytes of data. The storage cost is avoided with the employment of sixth-order, centred, finite differences schemes for $\mathbf{D}_x$ and $\mathbf{D}_y$, which improves the sparsity of $\mathbf{L}$ and lowers memory requirements from Gigabytes to Megabytes.


Other operators are constructed as follows:

\begin{equation}\label{eq:other_ops1}
	\mathbf{\Omega} = -i \omega \left[
	\begin{array}{cccc}
		\mathbf{I} & \mathbf{Z} & \mathbf{Z} & \mathbf{Z} \\
		\mathbf{Z} & \mathbf{I} & \mathbf{Z} & \mathbf{Z} \\
		\mathbf{Z} & \mathbf{Z} & \mathbf{I} & \mathbf{Z} \\
		\mathbf{Z} & \mathbf{Z} & \mathbf{Z} & \mathbf{Z}
	\end{array}
	\right], \quad
	\mathbf{H} = \left[
	\begin{array}{cccc}
		\mathbf{I} & \mathbf{Z} & \mathbf{Z} & \mathbf{Z} \\
		\mathbf{Z} & \mathbf{I} & \mathbf{Z} & \mathbf{Z} \\
		\mathbf{Z} & \mathbf{Z} & \mathbf{I} & \mathbf{Z}
	\end{array}
	\right]
\end{equation}

\begin{equation}\label{eq:other_ops2}
	\mathbf{B_u} = \left[
	\begin{array}{ccc}
		\mathbf{\sigma}^T \mathbf{I} & \mathbf{Z} & \mathbf{Z} \\
		\mathbf{Z} & \mathbf{\sigma}^T \mathbf{I} & \mathbf{Z} \\
		\mathbf{Z} & \mathbf{Z} & \mathbf{\sigma}^T \mathbf{I} \\
		\mathbf{Z} & \mathbf{Z} & \mathbf{Z}
	\end{array}
	\right], \quad
	\mathbf{B_f} = \left[
	\begin{array}{ccc}
		\mathbf{I} & \mathbf{Z} & \mathbf{Z} \\
		\mathbf{Z} & \mathbf{I} & \mathbf{Z} \\
		\mathbf{Z} & \mathbf{Z} & \mathbf{I} \\
		\mathbf{Z} & \mathbf{Z} & \mathbf{Z}
	\end{array}
	\right]
\end{equation}

Boundary conditions are inserted in $\mathbf{L}$, $\mathbf{B_u}$ and $\mathbf{B_f}$ by substituting the momentum equations in lines corresponding to positions at the boundaries. In other words, for the lower wall, eq. (\ref{eq:noslip}) gives
\begin{equation}\label{eq:bc_lower}
	\mathbf{L}^i = \mathbf{B_u}^i = \mathbf{B_f}^i = 0, \quad \forall i \: : \: y = 0
\end{equation}
and, for the upper limit,
\begin{equation} \label{eq:bc_upper}
	\frac{\partial}{\partial y} \mathbf{U}_{BL}(x,60) \approx 0
	\implies
	\left.
	\begin{array}{l}
		\mathbf{L}^i =
		\left[\begin{array}{cccc}
			\mathbf{D}_y & \mathbf{Z} & \mathbf{Z} & \mathbf{Z} \\
			\mathbf{Z} & \mathbf{D}_y & \mathbf{Z} & \mathbf{Z} \\
			\mathbf{Z} & \mathbf{Z} & \mathbf{D}_y & \mathbf{Z} \\
			\mathbf{Z} & \mathbf{Z} & \mathbf{Z} & \mathbf{Z}
		\end{array}\right]^i
		\\
		\mathbf{B_u}^i = \mathbf{B_f}^i = 0
	\end{array}
	\right\} ,\:
	\forall i \: : \: y = 60,
\end{equation}
according to eq. (\ref{eq:newmann_bc}). The superscript $\{\cdot\}^i$ refers to the $i$-th line of the corresponding matrix.

\section{Comparison between LES and reconstructed statistics} \label{sec:compPSD}

\begin{figure}
	\centering
	\begin{subfigure}[b]{.49\textwidth}
		\centering
		\includegraphics[width=\linewidth]{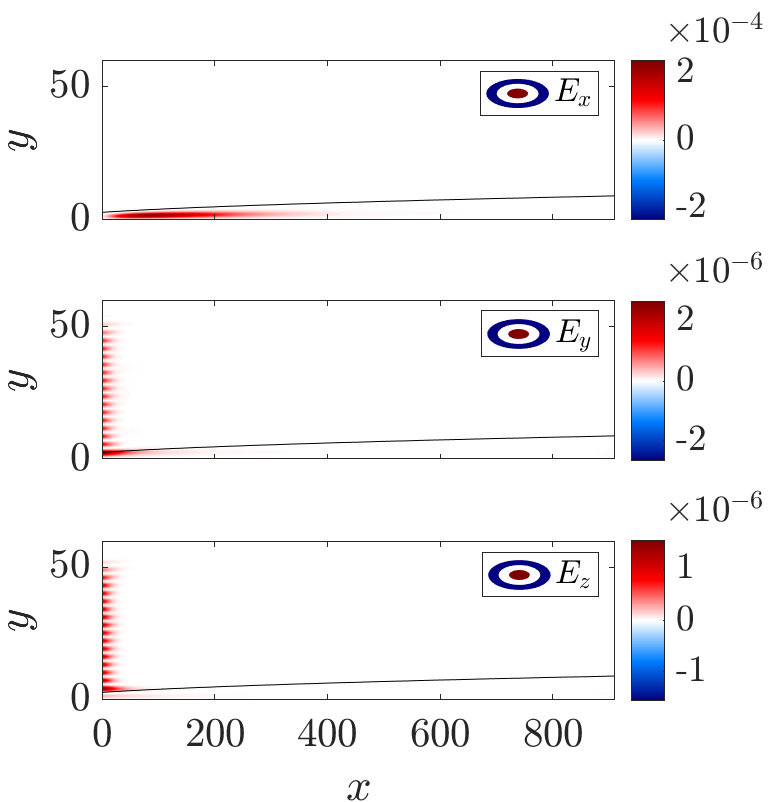}
		\caption{Statistical PSD, $\mathbf{P_U}$}
		\label{fig:PSD_Tu35_omega0.0000_beta1.1310}
	\end{subfigure}%
	\begin{subfigure}[b]{.49\textwidth}
		\centering
		\includegraphics[width=\linewidth]{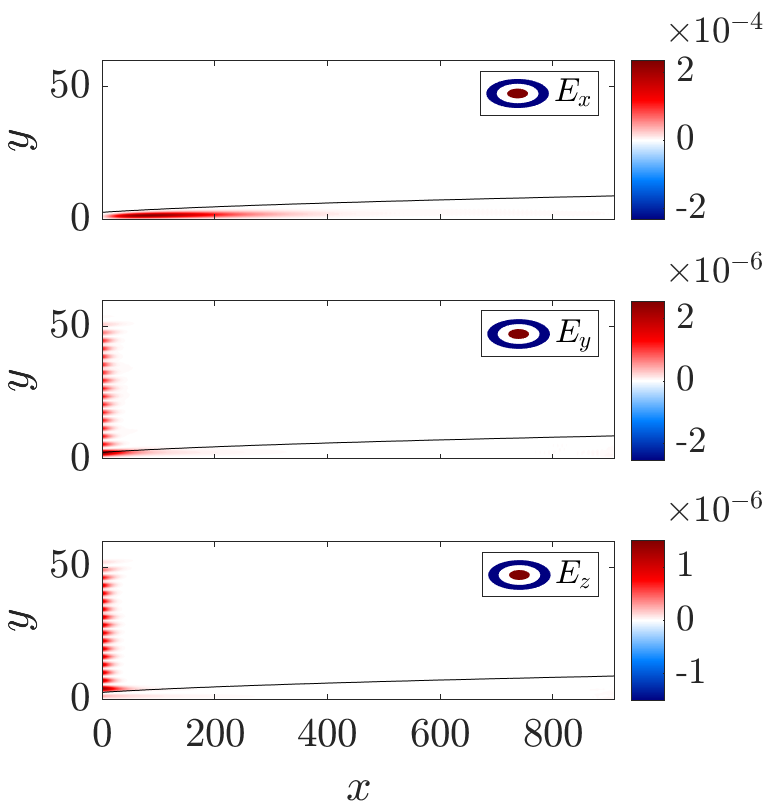}
		\caption{Reconstructed PSD, $\mathbf{P_Y}$}
		\label{fig:PSDrec_Tu35_omega0.0000_beta1.1310}
	\end{subfigure}
	\caption{Computed PSDs at $(\beta,\omega)=(1.131,0.000)$ and $Tu=3.5\%$.}
	\label{fig:PSD_linstreak}
\end{figure}

\begin{figure}
	\centering
	\includegraphics[width=0.9\linewidth]{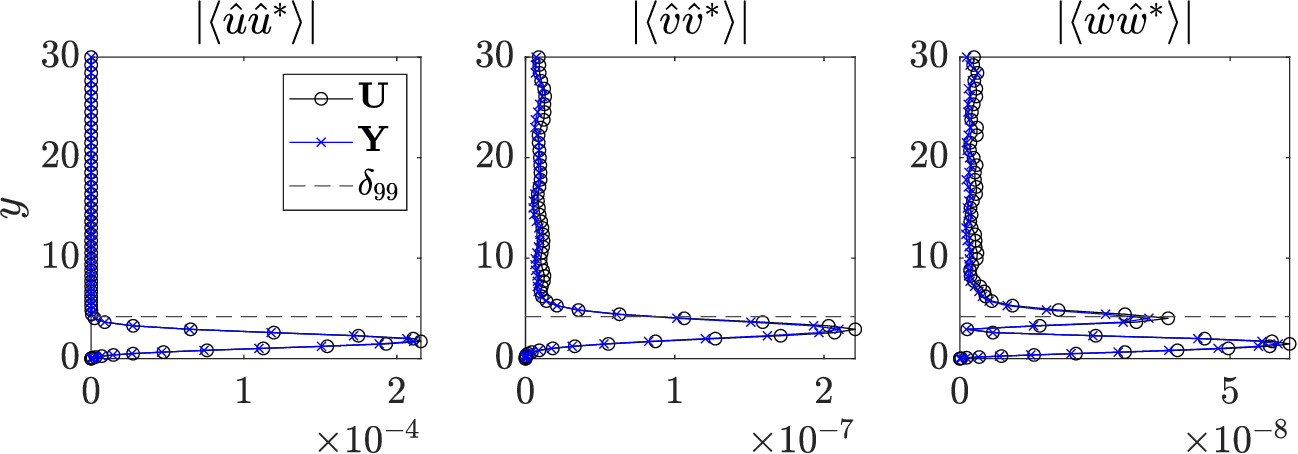}
	\caption{\revThree{Comparison between mean-squared velocities computed from LES statistics, $\mathbf{U}$, and reconstructed statistics, $\mathbf{Y}$, at $(\beta,\omega)=(1.131,0.000)$, $Tu=3.5\%$ and $x=113$. Respectively, $\langle \cdot \rangle$, $| \cdot |$ and $\{\cdot\}^*$ are average over blocks, absolute value and conjugate.}}
	\label{fig:supPSD_linstreak}
\end{figure}

\begin{figure}
	\centering
	\includegraphics[width=0.9\linewidth]{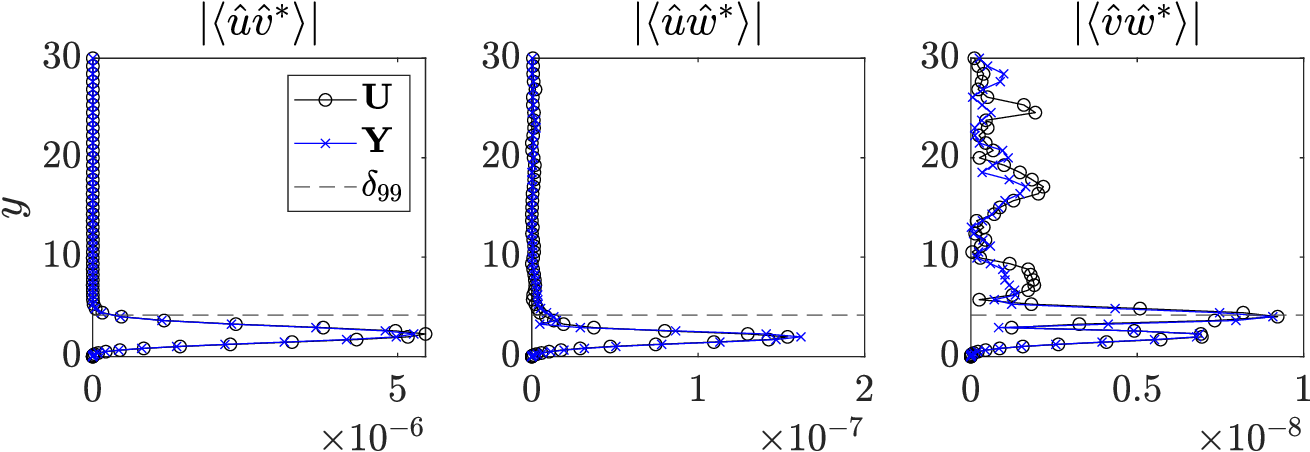}
	\caption{\revThree{Comparison between cross terms computed from LES statistics, $\mathbf{U}$, and reconstructed statistics, $\mathbf{Y}$, at $(\beta,\omega)=(1.131,0.000)$, $Tu=3.5\%$ and $x=113$. Respectively, $\langle \cdot \rangle$, $| \cdot |$ and $\{\cdot\}^*$ are average over blocks, absolute value and conjugate.}}
	\label{fig:supCross_linstreak}
\end{figure}

\begin{figure}
	\centering
	\begin{subfigure}[b]{.49\textwidth}
	\centering
	\includegraphics[width=\linewidth]{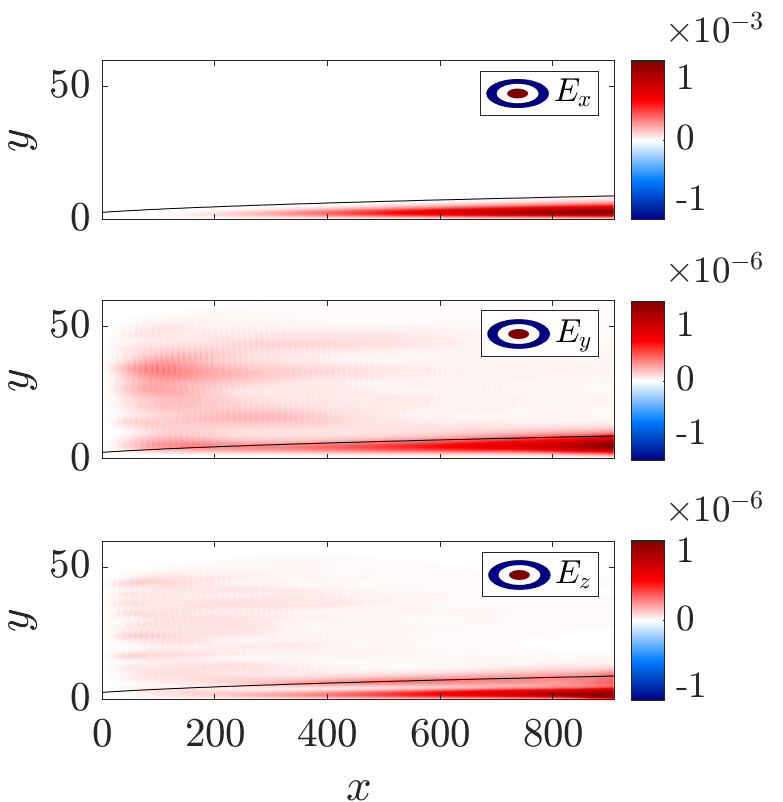}
	\caption{Statistical PSD, $\mathbf{P_U}$}
		\label{fig:PSD_Tu35_omega-0.0033_beta0.3770}
	\end{subfigure}%
	\begin{subfigure}[b]{.49\textwidth}
		\centering
		\includegraphics[width=\linewidth]{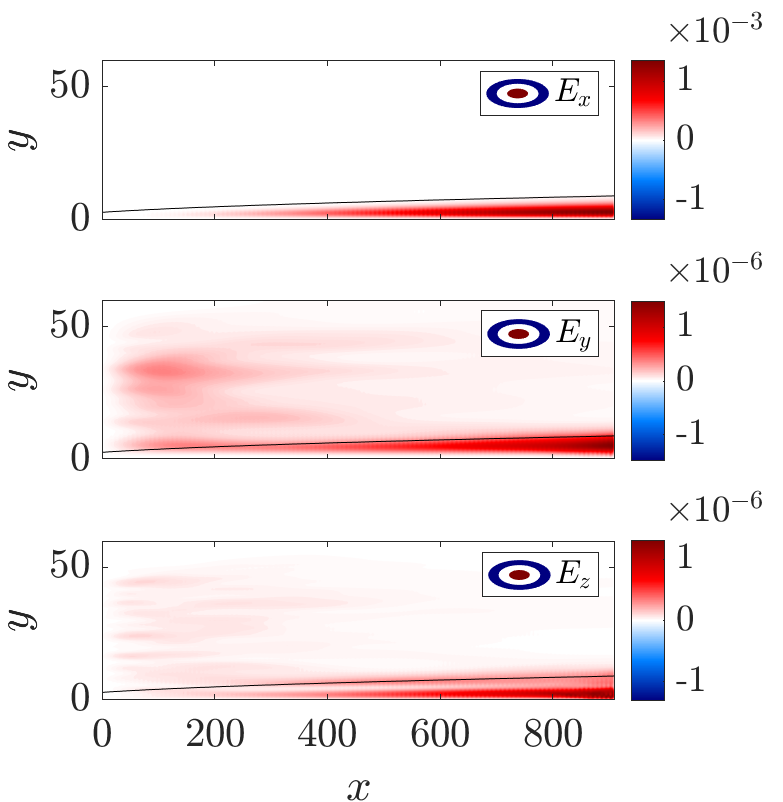}
		\caption{Reconstructed PSD, $\mathbf{P_Y}$}
		\label{fig:PSDrec_Tu35_omega-0.0033_beta0.3770}
	\end{subfigure}
	\caption{Computed PSDs at $(\beta,\omega)=(0.377,-0.003)$ and $Tu=3.5\%$.}
	\label{fig:PSD_nonlinstreak}
\end{figure}

\begin{figure}
	\centering
	\includegraphics[width=0.9\linewidth]{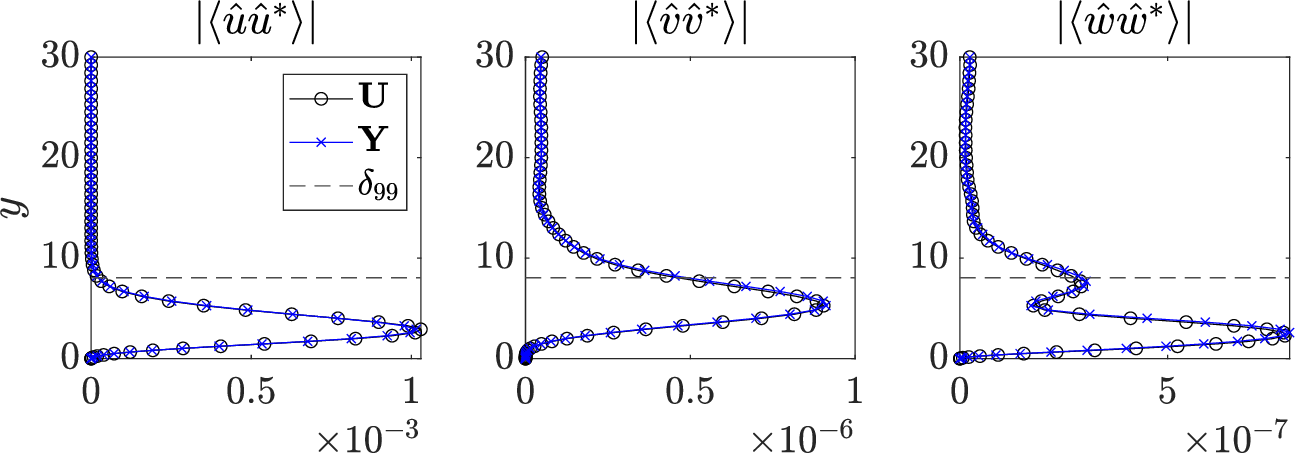}
	\caption{\revThree{Comparison between mean-squared velocities computed from LES statistics, $\mathbf{U}$, and reconstructed statistics, $\mathbf{Y}$, at $(\beta,\omega)=(0.377,-0.003)$, $Tu=3.5\%$ and $x=700$. Respectively, $\langle \cdot \rangle$, $| \cdot |$ and $\{\cdot\}^*$ are average over blocks, absolute value and conjugate.}}
	\label{fig:supPSD_nonlinstreak}
\end{figure}

\begin{figure}
	\centering
	\includegraphics[width=0.9\linewidth]{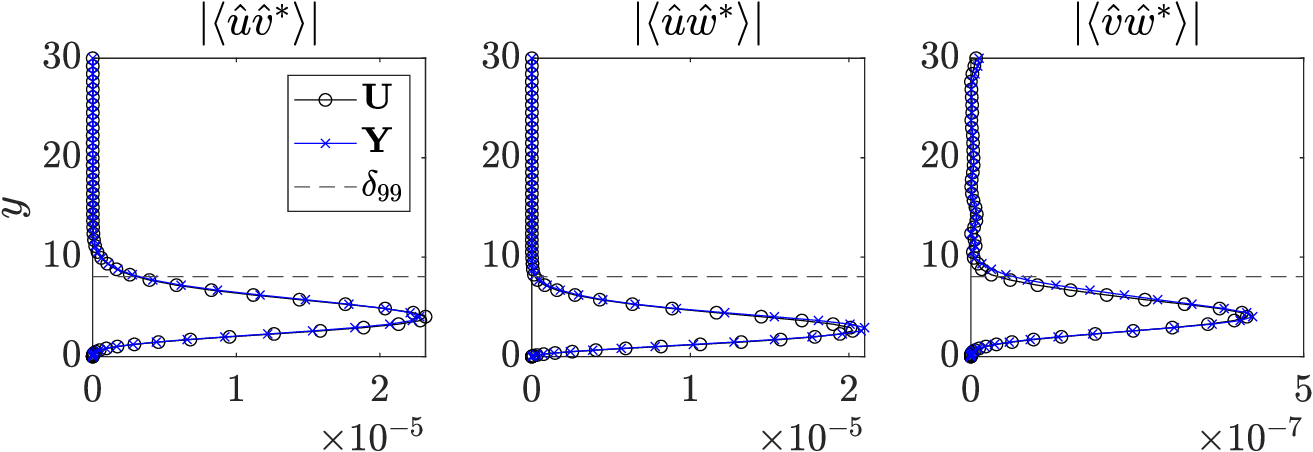}
	\caption{\revThree{Comparison between cross terms computed from LES statistics, $\mathbf{U}$, and reconstructed statistics, $\mathbf{Y}$, at $(\beta,\omega)=(0.377,-0.003)$, $Tu=3.5\%$ and $x=700$. Respectively, $\langle \cdot \rangle$, $| \cdot |$ and $\{\cdot\}^*$ are average over blocks, absolute value and conjugate.}}
	\label{fig:supCross_nonlinstreak}
\end{figure}

In the manuscript's text, only the linear and non-linear components of $\mathbf{P_Y}$, namely $\mathbf{\Pi_L}$ and $\mathbf{\Pi_N}$, were shown. For the sake of completeness, we display the computed statistics for the specific case of the streaks generated by the linear mechanism (figures \ref{fig:PSD_linstreak} to \ref{fig:supCross_linstreak}) and those resulting from non-linear mechanisms (figures \ref{fig:PSD_nonlinstreak} to \ref{fig:supCross_nonlinstreak}).

\bibliographystyle{abbrv}
\bibliography{refs.bib}

\end{document}